\documentclass{article}

\usepackage[utf8]{inputenc}
\usepackage{mathtools,csquotes}
\usepackage{amssymb}
\usepackage{geometry}
\usepackage{graphicx}
\usepackage{xcolor}
\usepackage{bm}
\usepackage{cite}
\usepackage{hyperref}

\hypersetup{
    colorlinks=true,
    linkcolor=blue,
    filecolor=magenta,      
    urlcolor=blue,
}
\newcommand{\w}{\omega}

\newcommand{\g}{\gamma}
\newcommand{\de}{\delta}

\newcommand{\dd}{\mathrm{d}}
\newcommand{\ii}{\mathrm{i}}
\newcommand{\dsR}{\mathbb{R}}
\newcommand{\scW}{\mathcal{W}}
\newcommand{\scL}{\mathcal{L}}

\title{On the role of geometric phase in the dynamics of elastic waveguides}

\author{
Mohit Kumar and Fabio Semperlotti\thanks{email: fsemperl@purdue.edu}
}

\date{Ray W. Herrick Laboratories, School of Mechanical Engineering, Purdue University, West Lafayette, Indiana 47907, USA\\[2ex]
\today}

\begin{document}

\maketitle



\begin{abstract}
    The geometric phase provides important mathematical insights to understand the fundamental nature and evolution of the dynamic response in a wide spectrum of systems ranging from quantum to classical mechanics. While the concept of geometric phase, which is an additional phase factor occurring in dynamical systems, holds the same meaning across different fields of application, its use and interpretation can acquire important nuances specific to the system of interest. In recent years, the development of quantum topological materials and its extension to classical mechanical systems have renewed the interest in the concept of geometric phase. This review revisits the concept of geometric phase and discusses, by means of either established or original results, its critical role in the design and dynamic behavior of elastic waveguides. Concepts of differential geometry and topology are put forward to provide a theoretical understanding of the geometric phase and its connection to the physical properties of the system. Then, the concept of geometric phase is applied to different types of elastic waveguides to explain how either topologically trivial or non-trivial behavior can emerge based on the geometric features of the waveguide.
\end{abstract}

\section{Introduction}
The term geometric phase indicates a phase factor acquired by a dynamical system following an adiabatic perturbation of the system parameters at the end of which the system is brought back to the original conditions. The geometric phase is in addition to the usual dynamical phase accumulated by a harmonically oscillating system. While first recognized in the context of quantum mechanics~\cite{berry1984quantal}, it was later observed also in other fields, such as optics~\cite{tomita1986observation,jisha2021geometric,cohen2019geometric}, solid state physics~\cite{zak1989berry,xiao2010berry,resta_macroscopic_1994}, classical mechanics~\cite{hannay1985angle}, and molecular physics~\cite{longuet1958studies,herzberg1963intersection}. The concept of geometric phase has been extended also to nonadiabatic~\cite{aharonov1987phase,anandan1992geometric} and nonlinear dissipative systems~\cite{ning1992geometric,kepler1992geometric,garrison1988complex}, its mathematical foundation has been firmly established~\cite{zwanziger1990berry,chruscinski_geometric_2004,simon1983holonomy}.

The geometric phase has proven to be a powerful concept to provide deep mathematical insight into certain aspects of dynamic phenomena. The ubiquity and usefulness of the geometric phase across several disciplines of wave physics and dynamics arise from its rigorous mathematical description and its interpretation via differential geometry and topology~\cite{simon1983holonomy}. For example, the geometric phase concept can describe the precession angle of a Foucault pendulum~\cite{von2007foucault}, provide an approach to sense the spatial distribution of scatterers in acoustic scattering problems~\cite{lata_2023_Underwater,lata2020topological}, model parameter variations required to realize non-Abelian physics~\cite{yang_2024_NonAbelian,chen2022classical}, and explain mathematical foundations and design principles underlying elastic topological metamaterials~\cite{hasan_colloquium_2010,bansil_colloquium_2016,chiu_classification_2016,xiao2015geometric,xin_topological_2020,huang2021recent,ma2019topological}. In this article, we will focus on the latter application as a way to provide a concrete example to illustrate the role of the geometric phase in the design of dynamical systems.

In recent years, topological metamaterials have gathered the interest of the physics and engineering communities. While the concept of topological metamaterial has found application in many areas of physics, in the following we focus on elastic topological metamaterials (ETMs) as a practical example to discuss the role of the geometric phase. Certainly many of these concepts are general and could be readily extended, for example, to photonics and phononics applications.
ETMs can be seen as a class of waveguides whose dynamic response can be understood, and even engineered, by using concepts of differential geometry and topology. As an example, over the past decade, ETMs have opened new ways to control and manipulate the propagation of stress waves. A very distinctive trademark of topological materials is the ability to support elastic waves (typically on boundaries or interfaces characterized by a topological transition) that are immune to back scattering from inhomogeneities and defects~\cite{liu_nonconventional_2019,liu_tunable_2018,liuSyntheticKramersPair2021}. The ETM field has grown extremely rapidly during the last decade also capitalizing on the many decades of research on (quantum) topological insulators~\cite{qi2011topological,hasan_colloquium_2010,chiu_classification_2016}, which first exploited the concept of topologically distinct classes of materials. The extension of these concepts to classical (i.e. non quantum mechanical) platforms, such as photonic~\cite{ozawa2019topological,lu2014topological} and acoustic systems~\cite{ma2019topological,xue2022topological}, also represented another key milestone to illustrate the feasibility of topological concepts for ETMs. Given that this review focuses on the role of geometric phase in the description of the dynamic behavior of elastic systems and uses ETMs only as an example of a potential application, a detailed discussion of ETMs is omitted. Nevertheless, the interested reader can refer to the following papers~\cite{yang2015topological,liu_nonconventional_2019,liu_tunable_2018,liuSyntheticKramersPair2021,susstrunk_classification_2016} and reviews~\cite{xue2022topological,xin_topological_2020,huang2021recent,miniaci2021design,ma2019topological,zhang_second_2023,shah_2024_Colloquium} as a starting point to explore the ETM field more in detail.

In view of the rising interest in using the geometric phase to study the dynamics of structures and materials, a cohesive discussion of its features and potential applications appears timely and necessary. This is not to say that the literature does not offer discussions on the geometric phase, but these discussions are often geared towards a quantum mechanical or theoretical physics audience~\cite{simon1983holonomy,zwanziger1990berry,chruscinski_geometric_2004,hasan_colloquium_2010}. The few discussions focused on applied mechanics tend to be scattered and limited to a narrow set of applications~\cite{ma2019topological,tromp1992berry,tromp1993surface,tromp1994surface,babich2005nongeometrical,segert1987photon,snieder2016seismic,budden1976phase,boulanger2012observation,deymier_geometric_2016,hasan_spectral_2019,deymier2017sound,lata_2023_Underwater,lata2020topological,hasan2022modeling}, therefore making approaching this topic and understanding the state of the art quite challenging.

In view of the above considerations, the present article aims to review the concept of geometric phase in the context of elastic waveguides and to leverage it to explain some fundamental ideas in ETMs.

The review is organized as follows. Section~\ref{sec:discrete} introduces the concept of geometric phase for systems described by a discrete set of variables. Section~\ref{sec:waves} reviews the formulation for elastic waveguides and provides examples of geometric phase. Section~\ref{sec:theory} introduces some relevant concepts from differential geometry that are applied in Section~\ref{sec:geometric-phase-holonomy} to explain the emergence of the geometric phase in elastic systems. Section~\ref{sec:applications} discusses the connection between the more theoretical mathematical ideas and the more practical dynamic behavior of elastic waveguides and ETMs. Section~\ref{sec:conclusions} summarizes the key points of this work.

\section{Geometric phase in discrete systems}
\label{sec:discrete}
The geometric phase appears in a wide range of oscillatory dynamical systems whose response depends on externally controlled time-varying parameters. Its appearance is well-studied in dynamical systems described by $n$ state variables, including discrete mechanical systems~\cite{hannay1985angle,deymier_geometric_2016}. Following the original formulation~\cite{berry1984quantal}, the governing equation can be written as 
\begin{equation}
\label{eqn:dynamical-system}
    \mathcal{D}_t\mathbf{x}+\scL(\mathbf{R}(t)) \mathbf{x}=0\;,
\end{equation}
where $\mathbf{x}$ is the vector of state variables, $\mathbf{R}(t)$ is the vector of controllable parameters, $t$ is time, $\mathcal{D}_t$ is a linear differential operator (typically $\mathcal{D}_t=d^2/dt^2$ in discrete mechanical systems, and $\mathcal{D}_t=\ii d/dt$ in quantum mechanical systems), and $\scL(\mathbf{R})$ is a matrix that depends on the parameters.

If the system's parameters are constant, the temporal evolution of the system is described by the modes of oscillation, which are derived by substituting the ansatz $\mathbf{x}(t)=\mathbf{X}e^{-\ii \lambda t}$ into Eq.~\eqref{eqn:dynamical-system}, which ultimately gives rise to an eigenvalue problem. The eigenvalue problem reads 
\begin{equation}
\label{eqn:dynamical-system-EVP}
    \scL(\mathbf{R}) \mathbf{X}(\mathbf{R}) = \lambda(\mathbf{R}) \mathbf{X}(\mathbf{R})\;,
\end{equation}
where the explicit dependence of the eigenvalues $\lambda$ and the eigenvectors $\mathbf{X}$ on the parameters is highlighted. Note that here $\mathbf{R}$ is constant by assumption If $\mathbf{X}$ is an eigenvector, then also $\mathbf{X} e^{\ii \g}$ is an eigenvector of the system, if $\g$ ranges between $0$ and $2\pi$. In this sense, the eigenvector has an \enquote{ambiguity} in phase. If two eigenvectors at a parameter value $\mathbf{R}_0$ have the same eigenvalue and differ by more than just a phase term, they are said to be degenerate. The parameter value $\mathbf{R}_0$ is said to be a degeneracy. In a general scenario, at least three parameters must be independently controlled to result in a degeneracy~\cite{berry1984quantal}. That is, degeneracies have codimension 3. These selected values of the parameters are referred to as \enquote{accidental degeneracies}.  However, if particular constraints (such as, for example, those arising from symmetries in the geometric configuration of the system) are present, then degeneracies will arise independently of the specific values of the parameters; or, equivalently, the degeneracies will not be lifted by simply choosing different values of the parameters (as long as the constraints are not violated). This latter class of degeneracies is said to be symmetry protected.

Suppose the parameters $\mathbf{R}(t)$ complete an adiabatic cycle in $T$ units of time. A cycle of the system parameters indicates that, in the parameter space, the point describing the properties of the system traces a closed loop $L$ (i.e., bringing the system back to the initial set of parameters). The adiabatic variation  dictates that the rate of change of parameters is \enquote{sufficiently slow}, hence implying that the time-scale of the parameters variation is much larger than the difference in the time-scale of oscillations of the excited modes (i.e., $T \gg \mathrm{max} \left( \frac{1}{\lambda_i - \lambda_j} \right)$, where $i,j=1,\dots,n$). Note that, in this context, the term \enquote{adiabatic} is used in its quantum mechanical sense (of a slowly-varying quantity). Therefore, it should not be interpreted in its thermodynamics sense, which instead indicates a \enquote{fast enough} process that prevents heat transfer between a system and its surroundings.

In this case, the evolution of a system is described by the adiabatic approximation~\cite{landau_mechanics_1982}. In particular, a system initialized in an eigenvector $\mathbf{X}(\mathbf{R}(0))$ evolves as 
\begin{equation}
\label{eqn:adiabatic-approx-discrete}
    \mathbf{x}(t) = \mathbf{X}(\mathbf{R}(t))  \exp (\ii \g(t)) \exp \left( -\ii \int_0^t \lambda(\mathbf{R}(t')) \, \dd t' \right)\;,
\end{equation}
where $\int_0^t \lambda(\mathbf{R}(t')) \, \dd t'$ is referred to as the dynamical phase~\cite{zwanziger1990berry} and $\g$ is referred to as the geometric phase. In other words, the state of a system at a time instant $t$ is determined by the mode of oscillation at the instantaneous parameter value $\mathbf{R}(t)$, the dynamical phase, and the geometric phase.
The dynamical phase describes the oscillatory nature of the system, for example, the phase acquired by a harmonic oscillator with a time-varying spring stiffness. If the parameters were constant, the dynamical phase would reduce to the familiar expression $\exp \left( -\ii \lambda(\mathbf{R}) t \right)$. The dynamical phase depends on the path traced in parameter space and the rate of parameter variation.

The geometric phase accounts for the phase ambiguity of the eigenvector. It is computed as~\cite{berry1984quantal} 
\begin{equation}
\label{eqn:geometric-phase-discrete}
    \g(t) = \int_{\mathbf{R}(0)}^{\mathbf{R}(t)} \mathrm{Im}\left[ \mathbf{X}(\mathbf{R})^*  \; \frac{\dd \mathbf{X}(\mathbf{R})}{\dd \mathbf{R}} \right] \dd \mathbf{R} \;,
\end{equation}
where $\mathbf{X}^*$ denotes the conjugate transposed. The derivation of Eq.~\eqref{eqn:geometric-phase-discrete} benefits from the mathematical discussions of Sec.~\ref{sec:theory} and it is postponed to Sec.~\ref{sec:geometric-phase-holonomy}. According to Eq.~\eqref{eqn:geometric-phase-discrete}, the geometric phase depends only on the path traced in parameter space, not on the rate of parameter variation, in contrast to the dynamical phase.

The geometric phase manifests most evidently at $t=T$, where the system properties match the initial values at $t=0$ ($\mathbf{R}(0)=\mathbf{R}(T)$). One may expect that the state of the system at $t=T$ would equal the state of the system at $t=0$ times the contribution of the dynamical phase. However, Eq.~\eqref{eqn:adiabatic-approx-discrete} shows that this would be the case only in the absence of the geometric phase, since
\begin{equation}
    \mathbf{x}(T) = \mathbf{X}(\mathbf{R}(0))  \exp (\ii \g(T)) \exp \left( -\ii \int_0^T \lambda(\mathbf{R}(t')) \, \dd t' \right)\;.
\end{equation}

The role of the geometric phase can be intuitively understood as being an intrinsic memory of the system~\cite{budden1976phase}. The phase acquired by an oscillatory system is, loosely speaking, the oscillator's \enquote{memory} because it records the oscillator's parameter variations over a given period of time (or, equivalently, the regions of the parameter space that the system has visited at a given time). This phase \enquote{memory} has two parts: the dynamical phase and the geometric phase. While the dynamical phase is the \enquote{memory} component familiar in classical (and quantum) physics, the geometric phase is an unexpected \enquote{additional memory}.


\section{Geometric phases in elastic waves}
\label{sec:waves}
The discussion of the occurrence and role of geometric phases in elastic continuous systems is somewhat scarce and scattered. To address this aspect and provide a more focused discussion of this concept from the perspective of elastic systems, we choose a prototypical system consisting in an isotropic quasi-one-dimensional (quasi-1D) waveguide subject to a cyclic and adiabatic variation of the cross section's geometry. By using this example, we will illustrate the type of mechanisms leading to the accumulation of geometric phase~\cite{tromp1992berry}. For quasi-1D waveguides, an adiabatic parameter variation means that the length-scale over which the parameters vary is much larger than the inverse of the difference in wavenumbers of propagating waves. The analysis can be generalized to include anisotropic and bulk materials~\cite{tromp1992berry,budden1976phase}. The geometric phase in the context of anisotropic bulk materials is discussed in supplementary material (Sec.~1))

There are two phase components available in elastic waves: the overall phase accumulated by a wave packet while traveling through the medium and the polarization angle, which is relevant for waveguides supporting polarized waves. The polarization angle specifies the \enquote{orientation} of the displacement field at a transverse plane of the waveguide, in a manner that will be further clarified in Sec.~\ref{sec:helical-waveguides}. A geometric phase can modify either one of the two phase components. 

The mathematical description of the geometric phase proceeds from the analysis of the Navier-Lam\'e (NL) equations of elastodynamics with traction-free boundary conditions. A quasi-1D waveguide whose longitudinal axis coincides with the $z$ axis is considered. The geometric properties of the cross section at $z$ are defined by the parameters $\mathbf{R}(z)$. For this waveguide, the NL equations and traction-free boundary conditions are~\cite{graffWaveMotionElastic2012}
\begin{gather}
    \label{eqn:NL}
    \mu \nabla^2 \mathbf{u} + (\mu + \lambda)\nabla(\nabla\cdot \mathbf{u}) = \rho \Ddot{\mathbf{u}}\;,\\
    \label{eqn:BCs}
    \sigma \cdot \Hat{n} = 0\;,
\end{gather}
where $\rho$ is the density, $\mathbf{u}$ is the displacement field, $\lambda$ and $\mu$ are the Lam\'e constants, $\sigma$ is the stress tensor, and $\Hat{n}$ is the outward unit normal at the surface. The stress tensor for an isotropic linear elastic solid is~\cite{sadd2009elasticity} $\sigma = C : \frac{1}{2}(\nabla \mathbf u + \nabla \mathbf{u}^\intercal)$, where $C$ is the fourth order stiffness tensor. 

If the parameters $\mathbf{R}$ are constant, the governing equations admit plane wave solutions of the form $\mathbf{u}(x,y,z,t)=\mathbf{U}(x,y)e^{\ii(k z - \w t)}$, where $\w$ is the angular frequency, $k$ is the wavenumber, and $\mathbf{U}(x,y)$ is the displacement field of a cross section. The term in the exponent $\phi_d = k z - \w t$ is the dynamical phase. These plane wave solutions are the guided modes of the waveguide that physically represent displacement field disturbances propagating along the longitudinal $z$ direction without changes in their cross-sectional displacement fields $\mathbf{U}(x,y)$. Guided modes are determined by substituting the plane wave ansatz into Eqs.~\eqref{eqn:NL},~\eqref{eqn:BCs}. This results in
\begin{gather}
    \label{eqn:eval}
    \mathcal{L}(k,\mathbf{R})\mathbf{U}(x,y;k,\mathbf{R})=\w^2 (k,\mathbf{R}) \mathbf{U}(x,y;k,\mathbf{R})\;,\\
    \label{eqn:eval-BC}
    \mathcal{B}(\mathbf{R})\mathbf{U}(x,y;k,\mathbf{R})=0\;,
\end{gather}
where $\mathcal{L}(k,\mathbf{R})$ and $\mathcal{B}(\mathbf{R})$ are linear differential operators. Equations~\eqref{eqn:eval},~\eqref{eqn:eval-BC} define an eigenvalue problem at each value of $k$. The eigenvalues define the so-called dispersion relations that link the angular frequency to the wavenumber of modes propagating in the waveguide. The eigenfunctions $\mathbf{U}(x,y;k,\mathbf{R})$ define the cross-sectional particle displacement fields associated with the guided modes. Further, the dependence of the dispersion relations and of the eigenfunctions on the parameter $\mathbf{R}$ is explicitly indicated in Eqs.~\eqref{eqn:eval},~\eqref{eqn:eval-BC}. If two guided modes at a parameter value $\mathbf{R}_0$ have identical eigenvalues for all values of $k$ (i.e. identical dispersion relations), the associated guided modes are said to be \enquote{fully degenerate}. The parameter value $\mathbf{R}_0$ will be called a \enquote{full degeneracy}. This choice of terminology is used to distinguish it from degeneracy in an eigenvalue problem (Sec.~\ref{sec:discrete}) that does not depend on a parameter such as $k$.

The general setup to study the geometric phase in the remainder of the paper considers a waveguide of length $Z$ with the geometric properties $\mathbf{R}$ varying adiabatically along the positive $z$ direction. Further, assume that the properties $\mathbf{R}(z)$ vary in a cycle between $z=0$ and $z=Z$, i.e., the properties at $z=0$ are identical to those at $z=Z$, written as $\mathbf{R}(0)=\mathbf{R}(Z)$. If the waveguide is harmonically excited with angular frequency $\w$ by the guided mode eigenfunction $\mathbf{U}(x,y;\mathbf{R}(0))$ at $z=0$, the steady state displacement field at $z$ is described by: (i) a guided mode eigenfunction $\mathbf{U}(x,y;\mathbf{R}(z))$, (ii) an amplitude factor $a(z)$ computed by the conservation of an adiabatic invariant, (iii) the usual dynamical component of phase $\phi_d$ computed as
\begin{equation}
\label{eqn:dyn_phase}
    \phi_d=\ii\int_0^z k(z') \dd z' - \ii \w t \;,
\end{equation}
and (iv) the geometric phase~\cite{tromp1992berry,bretherton1968propagation}. The computation of the geometric phase is deferred to Sec.~\ref{sec:geometric-phase-holonomy} as it can benefit from concepts of topology and differential geometry that will be introduced in Sec.~\ref{sec:theory}. 

As discussed in Sec.~\ref{sec:discrete}, the geometric phase appears as an additional phase component at $z=Z$ beyond the dynamical phase, representing an \enquote{additional memory} of the wave. The appearance of the geometric phase in elastic waveguides as a correction to the dynamical phase and polarization angle are illustrated in the following sections.

\subsection{Topological geometric phase: 1D waveguide with adiabatically-varying cross section}
\label{sec:1d-waveguide}
\begin{figure}
    \centering
    \includegraphics[width=\textwidth]{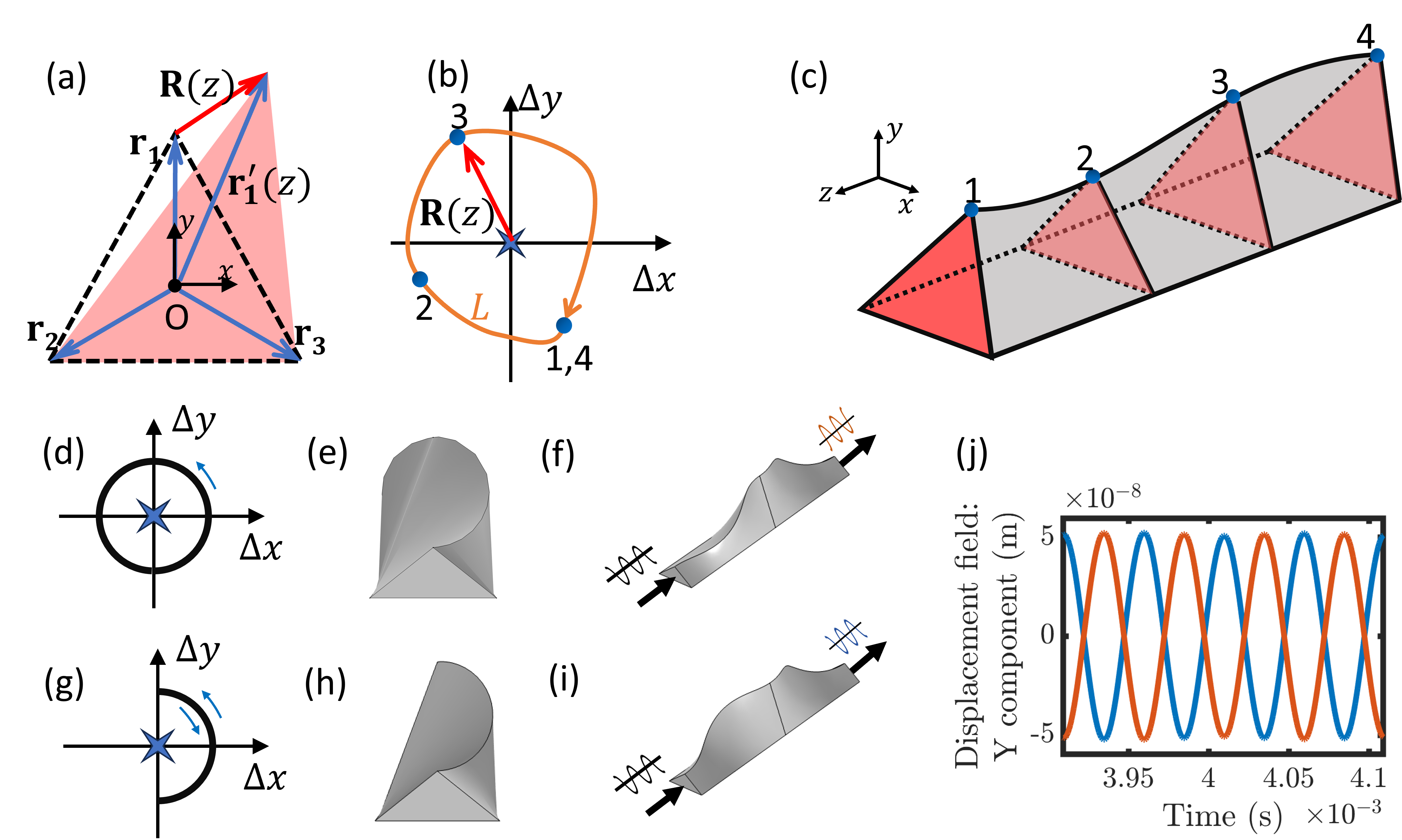}
    \caption{A topological geometric phase is observed in 1D waveguides with adiabatically varying cross sections along the longitudinal $z$ direction. (a) The shaded region is the cross section of the waveguide at a given $z$. The vertices $\mathbf{r}_1$, $\mathbf{r}_2$, and $\mathbf{r}_3$ form an equilateral triangle. $\mathbf{R}$ perturbs the cross section from an equilateral triangle. (b) The parameter space is comprised of the components of $\mathbf{R}$, namely, $\Delta x$ and $\Delta y$. The origin of the parameter space corresponds to an equilateral triangle, which corresponds to a full degeneracy. (c) A waveguide generated by tracing loop $L$ in the parameter space. The cross sections and points corresponding to the loop are marked 1-4. (d) A loop in the parameter space encircling the full degeneracy. (e,f) A front view and an orthographic view of the waveguide corresponding to (d). (g) A loop in the parameter space not encircling the full degeneracy. (h,i) A front view and an orthographic view of the waveguide corresponding to (g). Flexural guided modes are excited at the left end of the waveguide in (f) and (i), and the response is recorded at the right end. The longitudinal $z$ coordinate has been scaled down by a factor of $60$ in (f) and (i) for better visualization. (j) The $y$ component of the displacement field response at a point on the end of the waveguide is plotted for both waveguides.}
    \label{fig:pi-waveguide}
\end{figure}

This section presents an original example of geometric phase that emerges in a wave packet propagating in an elastic waveguide. Consider the case of flexural waves propagating in a 1D isotropic waveguide having an adiabatically-varying triangular cross section. A suitable parameter space to explore the role of adiabatic perturbations can be based on geometrical variables describing the cross section. A general triangular cross section can be described as a perturbation to an equilateral triangle. As shown in Fig.~\ref{fig:pi-waveguide}a, let the 2D coordinates of the vertices of an equilateral triangle be $\mathbf{r}_1$, $\mathbf{r}_2$, and $\mathbf{r}_3$ with respect to its centroid $O$. Consider a perturbation to $\mathbf{r}_1$ (i.e., the coordinate of the topmost vertex of the equilateral triangle) by the vector $\mathbf{R}=(\Delta x,\Delta y)$ to obtain $\mathbf{r}_1'=\mathbf{r}_1+\mathbf{R}$. The shaded triangle formed by the vertices with coordinates $\mathbf{r}'_1$, $\mathbf{r}_2$, and $\mathbf{r}_3$ represents a general cross section. 

The perturbations $(\Delta x,\Delta y)$ define a two-dimensional parameter space illustrated in Fig.~\ref{fig:pi-waveguide}b. The origin represents an equilateral triangular cross section. A uniform waveguide with this cross section has fully degenerate flexural modes that, in the long wavelength approximation, correspond to the in-plane and the out-of-plane bending modes. This point in parameter space (and its corresponding values) is a full degeneracy resulting from symmetry. Tracing out the loop $L$ in the parameter space of Fig.~\ref{fig:pi-waveguide}b leads to the waveguide in Fig.~\ref{fig:pi-waveguide}c whose cross section at the longitudinal coordinate $z$ (in the physical space) resembles Fig.~\ref{fig:pi-waveguide}a. 

One way to confirm the appearance of a geometric phase is to perform finite element simulations of the systems described above. Two stainless steel waveguides shown in Figs.~\ref{fig:pi-waveguide}f,i are generated by deforming an equilateral triangular cross section of side length 15~$mm$. In the parameter space, $\mathbf{R}(z)$ (i) traces a circle of radius 7~$mm$ centered at the origin (Fig.~\ref{fig:pi-waveguide}d) and (ii) traces forward and backward a semicircular arc of radius 7~$mm$ centered at the origin (Fig.~\ref{fig:pi-waveguide}g). The variation in cross section is implemented over a length of 6~$m$ to satisfy the adiabatic condition. One end of the waveguide is subject to a transverse harmonic excitation at a frequency $f=$~20~$kHz$ to produce a flexural mode. The other end is extended by 1.25~$m$ with a uniform cross section (not pictured) to prevent reflected waves from interfering with the desired measurements. The response in time is extracted at the right-hand termination of the waveguide by solving the model using the commercial finite element software COMSOL Multiphysics. Specifically, a transient analysis was performed using the solid mechanics interface.  The model was meshed with tetrahedral elements and by ensuring at least $N=$~8 elements per wavelength of the flexural mode. The time step of numerical integration was chosen to satisfy the Courant–Friedrichs–Lewy (CFL) condition~\cite{smith1985numerical}), $\Delta t = C_\mathrm{CFL}/fN$, where the CFL factor $C_\mathrm{CFL}$ was chosen as $0.1$.

Results show that the two waveguides produce time series that are perfectly out-of-phase with each other (Fig.~\ref{fig:pi-waveguide}j). This phase difference of $\pi$ is due to the geometric phase acquired by the waveguide of Fig.~\ref{fig:pi-waveguide}f. The choice of parameter variation causes both waveguides to accumulate the same amount of dynamical phase, so the difference in response is purely geometrical.

Next, two properties of this geometric phase are demonstrated with further numerical experiments. The first property is its independence from the rate of variation of the parameters as long as the adiabatic condition is satisfied. To observe this property, the same parameter variation of Figs.~\ref{fig:pi-waveguide}f,i is implemented over a length longer than 6~$m$ to generate the two waveguides. The dynamical phase associated with the response of each waveguide measured at the end (c.f. Eq.~\eqref{eqn:dyn_phase}) changes, but the two time series remain perfectly out of phase with each other. This result confirms that the acquired geometric phase remains $\pi$. The simulation results are presented in Sec.~2(a) of the supplementary material.

The second property of the geometric phase refers to the insensitivity to small perturbations of the parameter variations. For example, deforming the circular loop of Fig.~\ref{fig:pi-waveguide}d into an ellipse still results in a $\pi$ geometric phase (see Sec.~2(b) of the supplementary material for numerical simulations). As it will be discussed in more details in the following, the $\pi$ geometric phase appears for any loop in the parameter space encircling the origin (i.e., a full degeneracy). Further, this characteristic arises from the topology of the guided modes, and it motivates the classification of this geometric phase as \enquote{topological}. 

An additional novel example of a topological geometric phase in shear waves propagating in an anisotropic bulk material is presented in Sec.~1 of supplementary material.

\subsection{Nontopological geometric phase: Polarized waves in helical waveguides}
\label{sec:helical-waveguides}
\begin{figure}
    \centering
    \includegraphics[width=\linewidth]{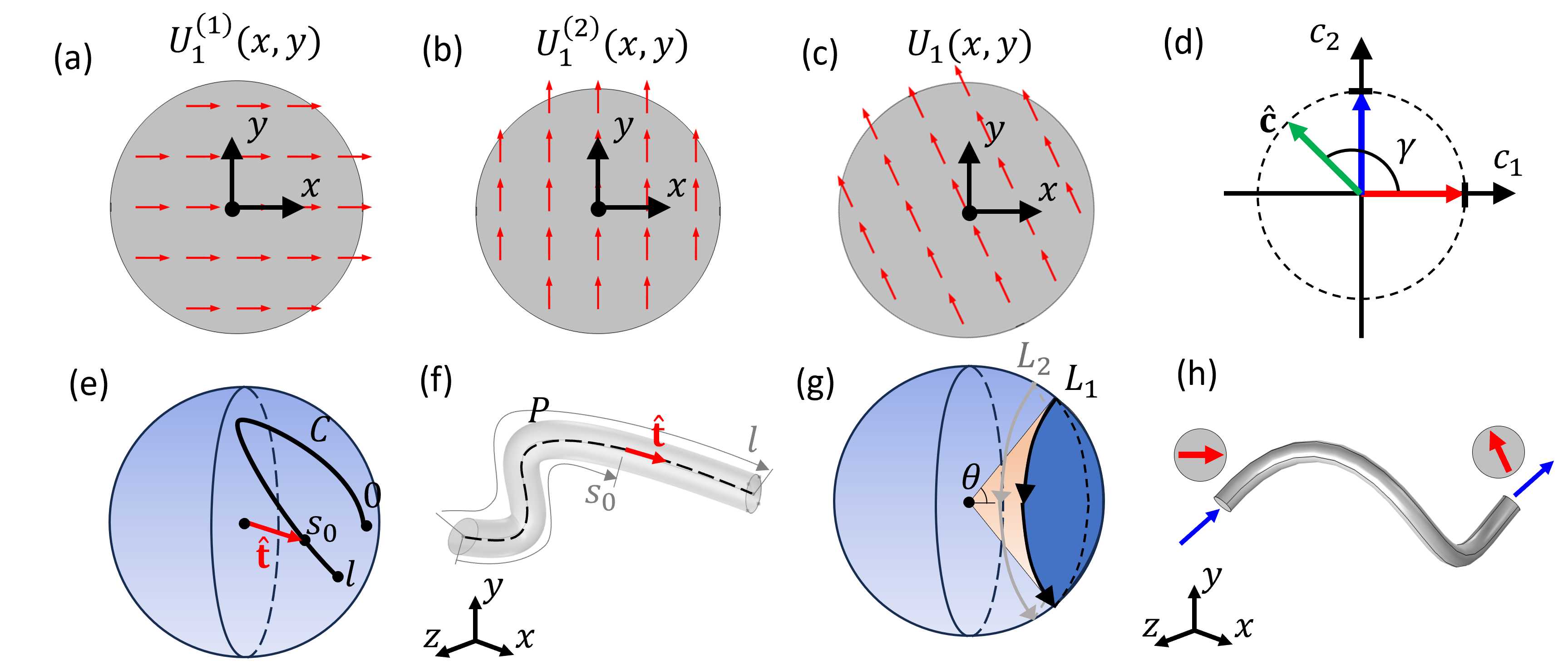}
    \caption{A nontopological geometric phase manifests itself in polarized waves propagating in a helical waveguide. The guided mode eigenfunction $\mathbf{U}$ of flexural guided modes in a uniform waveguide with circular cross section is plotted in (a)-(c). The out-of-plane component of the displacement is negligible. (a) A horizontally polarized ($x$-axis) guided mode. (b) A vertically polarized ($y$-axis) guided mode. (c) A guided mode with a general polarization. (d) Polarization vectors for (a) in red, (b) in blue, and (c) in green. The polarization angle of (c) is marked as $\g$. (e) A path $C$ on the spherical parameter space parameterized by $s$. $0$, $s_0$, and $l$ denote the path coordinate $s$ at the marked locations. (f) Path $P$ (dashed black line) created by the curve $C$ of (e) and the waveguide generated by sweeping a circular cross section along $P$. The unit tangent $\Hat{\mathbf{t}}(s_0)$ of path $P$ at a path coordinate $s_0$ corresponds to the point on the sphere $\Hat{\mathbf{t}}(s_0)$ on curve $C$ in (e). (g) Loops $L_1$ and $L_2$ in the parameter space. $\theta$ is the half-cone angle subtended by loop $L_1$. (h) A helical waveguide generated from curve $L_1$. A horizontally polarized flexural wave is excited at the left end. As it propagates down the waveguide, it acquires a polarization angle $\g$. The radius has been scaled up by a factor of 10 for better visualization.}
    \label{fig:polarization}
\end{figure}

This section explores an example of elastic waveguides that gives rise to a geometric phase of nontopological nature. Consider waveguides supporting a pair of fully degenerate flexural guided modes that have orthonormal eigenfunctions $\mathbf{U}^{(1)}(x,y)$ (Fig.~\ref{fig:polarization}a) and $\mathbf{U}^{(2)}(x,y)$ (Fig.~\ref{fig:polarization}b), which correspond to horizontal and vertical particle displacements respectively. Orthonormality implies $\int_S \mathbf{U}^{(i)*} \cdot \mathbf{U}^{(j)}\, \dd x \dd y = \de_{ij}$, where $S$ is the cross section, $\de_{ij}$ is the Kronecker delta, $i=1,2$, and $j=1,2$. A linear combination of these eigenfunctions, $\mathbf{U}=c_1 \mathbf{U}^{(1)}+ c_2 \mathbf{U}^{(2)}$, represents a guided mode with the particle displacements along the unit vector $\Hat{\mathbf{c}}=(c_1 \Hat{\mathbf{e}}_x + c_2 \Hat{\mathbf{e}}_y) / \sqrt{c_1^2 + c_2^2}$. (Fig.~\ref{fig:polarization}c). The plane wave solution corresponding to $\mathbf{U}$ is typically referred to as a polarized wave. The unit vector $\Hat{\mathbf{c}}$ is called the polarization vector (Fig.~\ref{fig:polarization}d). The angle subtended between $\Hat{\mathbf{c}}$ and the $c_1$ axis is the polarization angle $\g$ of the wave. A guided mode eigenfunction proportional to $\mathbf{U}^{(1)}$, shown in Fig.~\ref{fig:polarization}a, ($\mathbf{U}^{(2)}$, Fig.~\ref{fig:polarization}b) is said to be horizontally (vertically) polarized as its polarization vector marked by the red (blue) arrow is horizontal (vertical) in Fig.~\ref{fig:polarization}d.

Polarized waves provide several examples of nontopological geometric phases~\cite{boulanger2012observation,segert1987photon,snieder2016seismic}, which resemble the precession motion of a Foucault pendulum in terms of the associated parameter space and eigenfunctions~\cite{snieder2016seismic,von2007foucault}. Additional examples of nontopological geometric phases are provided by surface waves~\cite{tromp1993surface,tromp1994surface,babich2005nongeometrical} and waves in rotating media~\cite{budden1976phase,tromp1994surface}.

We illustrate the concept of nontopological geometric phase by means of helical waveguides created by sweeping a circular cross section through a generating path $P$ (in real space)~\cite{boulanger2012observation,segert1987photon}. Such waveguides support polarized flexural waves, as each cross section normal to the generating path is circular, and circular cross sections support fully degenerate flexural modes~\cite{graffWaveMotionElastic2012}. The geometry of a waveguide (see Fig.~\ref{fig:polarization}f) is specified by the radius $r$ of the cross section (fixed to be 0.05~$m$), the length $l$ of the centerline of the waveguide, and a curve $C$ on a spherical parameter space (Fig.~\ref{fig:polarization}e). $l$ determines the length of the generating path $P$ in Fig.~\ref{fig:polarization}f. Curve $C$ in Fig.~\ref{fig:polarization}e specifies the unit tangent vector $\Hat{\mathbf{t}}(s_0)$ of the generating path $P$ at the path coordinate $s_0$ (Fig.~\ref{fig:polarization}f). Together, $l$ and $C_1$ specify the generating path $P_1$ along which the circular cross section of radius $r$ is swept to generate the waveguide in Fig.~\ref{fig:polarization}f. $l$ is assumed large enough to ensure the adiabatic variation of the path tangents. In what follows, the curve on the unit sphere is assumed to be a closed loop, such as curve $L_1$ in Fig.~\ref{fig:polarization}g. This assumption implies that the initial and final tangent vectors of the generating path are parallel. Then, as a wave propagates along the waveguide, the direction of propagation given by $\Hat{\mathbf{t}}$ adiabatically completes a cycle. 

As in the previous section, we can show the occurrence and properties of the geometric phase in helical waveguides by exploiting numerical simulations. Consider the helical waveguide depicted in Fig.~\ref{fig:polarization}h generated by the loop $L_1$ in Fig.~\ref{fig:polarization}g. The generating helical path has a radius $r_h=$~2.4~$m$, pitch $p=$~14~$m$, and consists of one revolution. The length of the path is $l=\sqrt{(2\pi r_h)^2 + p^2}=$~20.58~$m$. The angle between the unit tangent vector and the axis of the helix denoted by $\theta$ is constant (Fig.~\ref{fig:polarization}g) and equals $\theta = \cos^{-1}(p/l)=47.12^\circ$~\cite{tomita1986observation}. The reference eigenfunctions $\mathbf{U}^{(1)}$ and $\mathbf{U}^{(2)}$ of two fully degenerate and orthonormal guided modes corresponding to horizontal and vertical polarizations are shown in Figs.~\ref{fig:polarization}a,b. One end of the waveguide is excited harmonically to produce a horizontally polarized flexural wave. The displacement field response is recorded at the other end of the waveguide. The recorded guided mode eigenfunction is plotted in Fig.~\ref{fig:polarization}c, indicating that the polarization angle does not return to zero at the end of the waveguide. The polarization angle is computed as $\g=114^\circ$ by projecting the displacement profile onto the horizontally and vertically polarized guided modes (Fig.~\ref{fig:polarization}d). This polarization angle is a geometric phase. As will be discussed in Sec.~\ref{sec:parallel-transport}, the geometric phase is analytically computed as the solid angle subtended by the loop $L_1$. The analytical result is $\g_{\mathrm{theory}}=-2 \pi\cos \theta \mod 2\pi =-244.9^\circ \mod 360^\circ=115.1^\circ$, which agrees well with the numerical computation.

By repeating the simulation with a greater value of the generating path length $l$, the geometric phase is verified to be independent from the rate of variation of the parameters (as long as the adiabatic condition is satisfied). However, the geometric phase is not topologically protected. Repeating the simulation after perturbing the curve $L_1$ to $L_2$ changes the polarization angle to $129^\circ$. The corresponding simulations are reported in Sec.~3 of supplementary material. Indeed, changing the loop in the spherical parameter space changes the solid angle it subtends, which in turn changes the geometric phase. This is why the geometric phase discussed here is classified as \enquote{nontopological}.

\subsection{Summary of the general characteristics of the geometric phase}
The defining characteristics of the geometric phase in elastic waveguides are summarized as follows. First, waves in elastic waveguides that adiabatically vary along the direction of wave propagation can exhibit geometric phases either as a correction to the dynamical phase or to the polarization angle. Second, a geometric phase depends solely on the path described by the parameters in the parameter space, and it is independent of the rate of variation of these parameters. This property justifies the terminology \enquote{geometric}. Third, geometric phases are categorized into two classes: topological and nontopological. Topological geometric phases are associated with full degeneracies in the associated eigenvalue problem describing the dynamics of the system (Eqs.~\eqref{eqn:eval},~\eqref{eqn:eval-BC}). Physically, if a waveguide exhibits a topological geometric phase, the value of the geometric phase will not be affected by small perturbations to either the geometry or the composition (e.g. the material distribution) of the waveguide. On the other hand, if a waveguide exhibits a nontopological geometric phase, the value of the geometric phase will change with perturbations to the properties of the waveguide.

\section{Concepts from differential geometry}
\label{sec:theory}
The geometric phase can be further described and understood by using differential geometry~\cite{nakahara2003geometry}. Two concepts are introduced: fiber bundles and parallel transport. 

\subsection{Fiber bundles}
\label{sec:fiber-bundles}
\begin{figure}
    \centering
    \includegraphics[width=\linewidth]{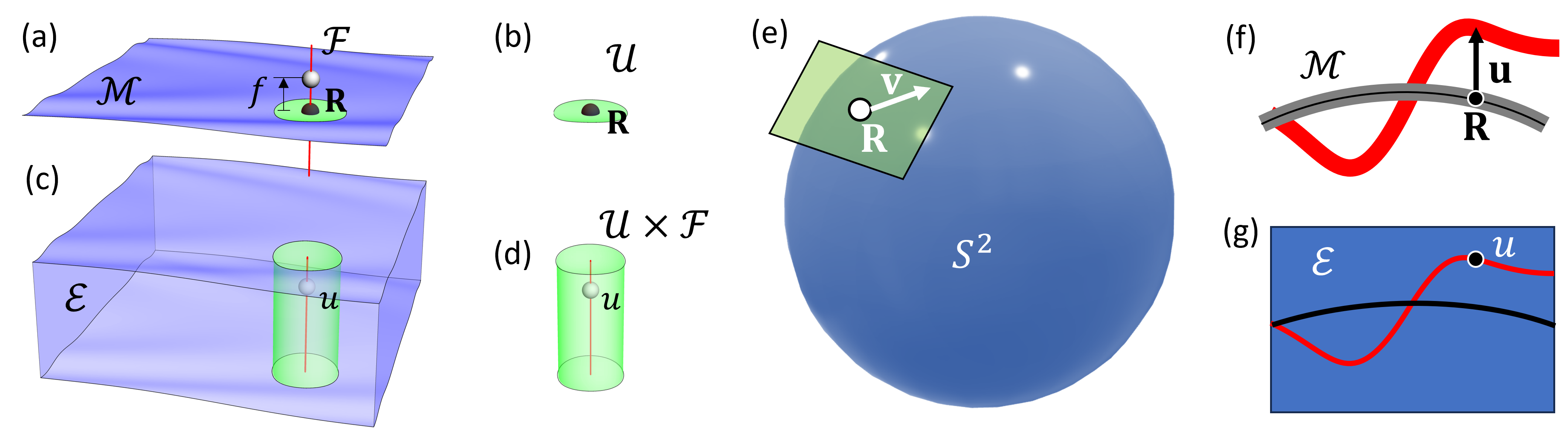}
    \caption{(a) A two-dimensional manifold $\mathcal{M}$. At a point $\mathbf{R}$, a 1D fiber $\mathcal{F}$ can be attached to create the fiber bundle in (c). (b) A point $\mathbf{R}$ of the manifold $\mathcal{M}$ is in a neighborhood $\mathcal{U}$ that is equivalent to the Euclidean plane $\dsR^2$. (c) A fiber bundle $\mathcal{E}$ created from a two-dimensional base manifold $\mathcal{M}$ and a 1D fiber $\mathcal{F}$. The pictured fiber bundle is trivial. The marked point $u$ has a base coordinate $\mathbf{R}$ and fiber coordinate $f$. (d) $u$ lies in the neighborhood $\mathcal{U}\times\mathcal{F}$.  (e) A sphere with a tangent vector $\mathbf{v}$ attached at the point $\mathbf{R}$. The green plane is a tangent plane attached at $\mathbf{R}$. Attaching the tangent plane at each point gives rise to the tangent bundle of a sphere $TS^2$. (f) A curved beam in undeformed (gray) and deformed (red) configurations. (g) Fiber bundle description of the curved beam modeled by the Euler-Bernoulli theory. The base manifold $\mathcal{M}$ is the centerline of the undeformed beam, and the fiber is $\dsR$ to model the transverse displacement of the beam. They lead to the (trivial) fiber bundle $\mathcal{E} = \mathcal{M} \times \dsR$. The marked point $u$ has the base coordinate $\mathbf{R}$ and fiber coordinate $\mathbf{u}$. Sections of the fiber bundle corresponding to the undeformed (black) and deformed (red) beam are marked.}
    \label{fig:manifold-FB}
\end{figure}

A $n$-dimensional \textit{manifold} $\mathcal{M}$ is a space that is locally equivalent to the Euclidean space $\dsR^n$ (Fig.~\ref{fig:manifold-FB}a). That is, each point $\mathbf{R}$ has a neighborhood $\mathcal{U}$ that is equivalent to the Euclidean space (Fig.~\ref{fig:manifold-FB}b). Two manifolds are said to be topologically equivalent, or simply equivalent, if they can be continuously deformed into one another. The entire manifold is, in general, topologically distinct (i.e., not equivalent) from a Euclidean space because its constituent neighborhoods may be \enquote{glued} together in ways not replicable in a Euclidean space. Familiar manifolds used in geometry are the plane $\dsR^2$, the cylinder, the sphere, and the torus. The latter three manifolds are not globally equivalent to the plane.

At each point $\mathbf{R}$ of a $n$-dimensional manifold $\mathcal{M}$, another $m$-dimensional manifold $\mathcal{F}$ may be attached (Fig.~\ref{fig:manifold-FB}a). This construction gives rise to a $(n+m)$-dimensional manifold, called a \textit{fiber bundle} $\mathcal{E}$, which is illustrated in Fig.~\ref{fig:manifold-FB}c. A point $u$ of the fiber bundle contains information of a point $\mathbf{R}$ of the base manifold and a point $f$ on the fiber at $\mathbf{R}$. In this context, $\mathcal{M}$ is the \textit{base manifold}, $\mathbf{R}$ is a \textit{base coordinate}, $\mathcal{F}$ is the \textit{fiber}, and $f$ is a \textit{fiber element} or \textit{fiber coordinate}. A fiber bundle is locally equivalent to a Cartesian product of manifolds: a point $u$ on the fiber bundle has a neighborhood $\mathcal{U}\times\mathcal{F}$ (Fig.~\ref{fig:manifold-FB}d), where $\mathcal{U}$ is a neighborhood of the base manifold containing the base coordinate of $u$. However, the fiber bundle $\mathcal{E}$ is, in general, topologically distinct from the Cartesian product of manifolds $\mathcal{M} \times \mathcal{F}$ because the constituent neighborhoods may be \enquote{glued} together in ways not replicable by a Cartesian product. 

If the fiber bundle $\mathcal{E}$ is equivalent to $\mathcal{M} \times \mathcal{F}$, it is said to be trivial. The trivial nature of a fiber bundle will prove important in subsequent discussions on the topological geometric phase. This property can be observed from the following mathematical result (Theorem 9.1 of Ref.~\cite{nakahara2003geometry}): ``A fiber bundle defined over a base manifold equivalent to a Euclidean space is trivial.''

The tangent bundle $T\mathcal{M}$ is a familiar example of a fiber bundle, where the fiber is the tangent space of the base manifold $\mathcal{M}$ and a fiber element is a tangent vector. The fiber bundle is constructed by attaching a tangent space to each point of the base manifold. Considering the example of a sphere, this implies attaching tangent planes to each point on the sphere, as illustrated in Fig.~\ref{fig:manifold-FB}e. The tangent bundle of a sphere is four-dimensional.

\begin{figure}
    \centering
    \includegraphics[width=\linewidth]{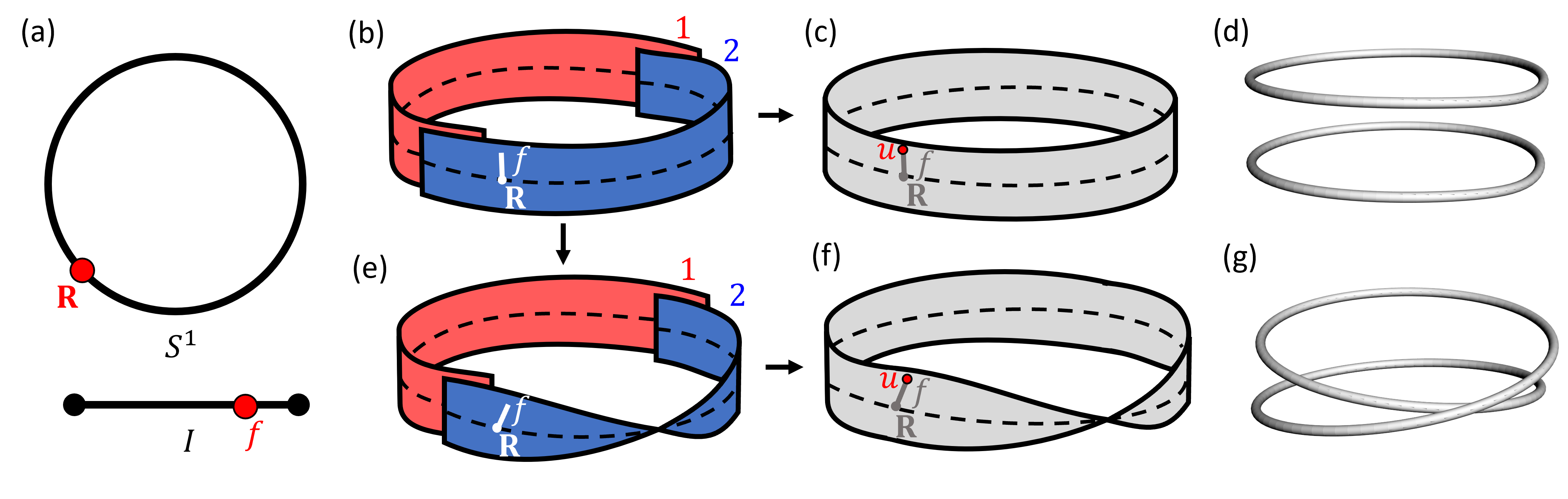}
    \caption{Constructing a cylinder and a M\"obius strip as fiber bundles. (a) The base manifold $S^1$ and fiber $I=[-1,1]$. (b) Strips 1 and 2 are the pieces of the fiber bundle. (c) The fiber bundle obtained by gluing the strips 1 and 2 end-to-end is a cylinder. (d) The edges of a cylinder. (e) One of the strips may be given a half-twist before gluing. (f) The fiber bundle obtained by gluing the arrangement in (d) is a M\"obius strip. (g) The edge of a M\"obius strip.}
    \label{fig:cylinder_mobius}
\end{figure}

Two fiber bundles of relevance for topological geometric phases are the cylinder and the M\"obius strip. Both are constructed from a circular base manifold $S^1$ and fiber $I=[-1,1]$ (Fig.~\ref{fig:cylinder_mobius}a). The difference lies in the connections between their component pieces (strips 1 and 2 in Figs.~\ref{fig:cylinder_mobius}b,e). If the two strips are joined end-to-end (Fig.~\ref{fig:cylinder_mobius}b), the construction results in a cylinder (Fig.~\ref{fig:cylinder_mobius}c). The cylinder is a trivial fiber bundle as it is equivalent to $S^1 \times I$. If one of the strips is given a half-twist prior to gluing (Fig.~\ref{fig:cylinder_mobius}e), the construction results in a M\"obius strip (Fig.~\ref{fig:cylinder_mobius}f). The M\"obius strip is a nontrivial fiber bundle. The cylinder and the M\"obius strip are topologically distinct as they cannot be continuously deformed into one another. During the discussion of topological geometric phases, the fiber in the above construction will be a two point set $I=\{0,\pi\}$. The resulting fiber bundles correspond to the \enquote{edges} of the cylinder (Fig.~\ref{fig:cylinder_mobius}d) and the M\"obius strip (Fig.~\ref{fig:cylinder_mobius}g).

The physical relevance of the fiber bundles lies in their ability to capture the response of position- or parameter-dependent systems. The space of allowable values of either positions or parameters gives rise to the base manifold. At each position or parameter, the space of possible responses of the system is modeled as a fiber. An example of a position-dependent system modeled as a fiber bundle is the deformation of an elastic body (Fig.~\ref{fig:manifold-FB}f,g). An undeformed elastic body defines a manifold $\mathcal{M}$ (Fig.~\ref{fig:manifold-FB}f). Each point of the elastic body is associated with a displacement vector, which can be modeled as an element of a fiber. The fiber depends on the permissible displacements: for example, a general 3D problem requires $\dsR^3$ as the fiber, a plane strain problem requires $\dsR^2$ as the fiber, and approximate 1D theories, such as the Euler-Bernoulli beam theory, require $\dsR$ as the fiber. A point $u$ on the resulting fiber bundle $\mathcal{E}$ has a base coordinate $\mathbf{R}$, which describes a point on the elastic body, and a fiber coordinate $\mathbf{u}$, which describes the displacement at $\mathbf{R}$ (Fig.~\ref{fig:manifold-FB}f,g). A deformed configuration of the body specifies $\mathbf{u}(\mathbf{R})$, which corresponds to a \textit{section} of the fiber bundle. Parameter-dependent systems modeled as fiber bundles form the basis to understand the geometric phase (Sec.~\ref{sec:geometric-phase-holonomy}).

\subsection{Connections and parallel transport}
\label{sec:parallel-transport}
The physical perspective of fiber bundles modeling position- or parameter-dependent systems suggests that varying the base coordinate (e.g., position on an elastic body) alters the fiber coordinate (e.g., displacement vector at that position). The process of varying an element of the fiber bundle by moving the base coordinate along a curve on the base manifold and recording the associated change in the fiber coordinate is known as \textit{parallel transport}. The rule relating the change in fiber coordinates to the change in base coordinates is known as a \textit{connection}. 

\begin{figure}
    \centering
    \includegraphics[width=0.75\linewidth]{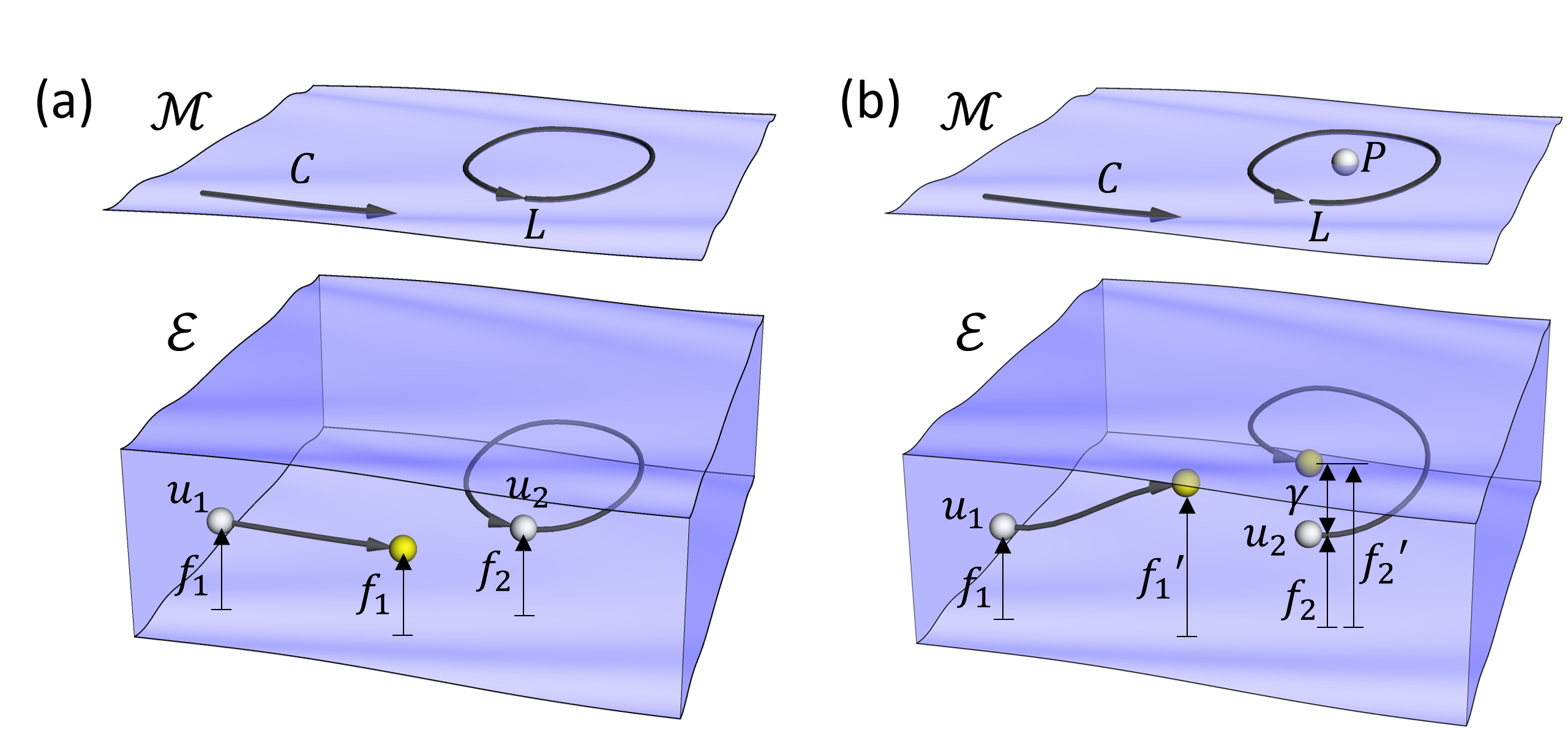}
    \caption{Parallel transport on the fiber bundle of Fig.~\ref{fig:manifold-FB} according to a connection. $C$ and $L$ are a curve and a loop on the base manifold, respectively. (a) Fiber bundle with a flat connection. Parallel transport of an element $u_1$ along $C$ does not change the fiber coordinate $f_1$. It follows that parallel transport of an element $u_2$ along the loop $L$ does not change it. (b) Fiber bundle with a curved connection. Parallel transport of an element $u_1$ along $C$ changes the fiber coordinate from $f_1$ to $f_1'$. Parallel transport of an element $u_2$ along the loop $L$ changes the fiber coordinate from $f_2$ to $f_2'$, resulting in a holonomy of $\g$.}
    \label{fig:connection}
\end{figure}

Figure~\ref{fig:connection} illustrates parallel transport on the fiber bundle of Fig.~\ref{fig:manifold-FB} using two (contrived) connections. The first connection does not change the fiber coordinate during the parallel transport of an element of the fiber bundle (Fig.~\ref{fig:connection}a). Such a connection is said to be \textit{flat}. The second connection, in general, changes the fiber coordinate during the parallel transport of an element of the fiber bundle (Fig.~\ref{fig:connection}b). Such a connection is said to be \textit{curved}. In particular, the connection in Fig.~\ref{fig:connection}b increases the fiber coordinate in proportion to the unwrapped winding angle of a path about point $P$. The change in fiber coordinate after parallel transport of an element of the fiber bundle around a loop in the base manifold is known as \textit{holonomy} ($\g$ in Fig.~\ref{fig:connection}b).

\begin{figure}
    \centering
    \includegraphics[width=\linewidth]{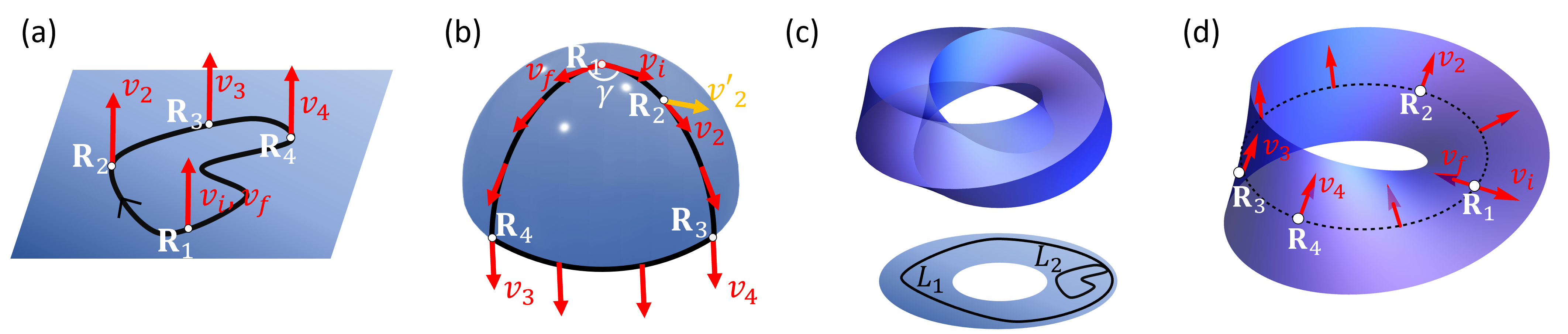}
    \caption{Parallel transport on fiber bundles. The fiber element $v_i$ at $\mathbf{R}_1$ transforms to $v_f$ after parallel transport along the loop $\mathbf{R}_1 \mathbf{R}_2 \mathbf{R}_3 \mathbf{R}_4$. (a) Parallel transport on the normal bundle $\mathcal{N}\dsR^2$. (b) Parallel transport on the tangent bundle of a hemisphere. $v_i$ at $\mathbf{R}_1$ under parallel transport turns into $v_2$ at $\mathbf{R}_2$. It does not turn into to $v_2^\prime$ at $\mathbf{R}_2$ because the connection prevents rotation about the local normal. (c) A thickened M\"obius strip and its annular base manifold. $L_1$ and $L_2$ are two loops on the base manifold. (d) The thickened M\"obius strip restricted to the loop $L_1$ is a M\"obius strip.}
    \label{fig:parallel-transport}
\end{figure}

The appearance of a holonomy depends on the nature of the fiber bundle (either trivial or nontrivial) and on the curvature of the connection (either flat or curved). Three representative examples from geometry are discussed in Fig.~\ref{fig:parallel-transport}.

Figure~\ref{fig:parallel-transport}a illustrates parallel transport on a bundle constructed by attaching a perpendicular line $\dsR$ at each point on the plane $\dsR^2$. The bundle is known as the normal bundle over the plane, and it is denoted as $\mathcal{N}\dsR^2$. Its base manifold is $\dsR^2$, and the fiber is $\dsR$. As the base manifold is a Euclidean space, $\mathcal{N}\dsR^2$ is trivial. $\mathcal{N}\dsR^2$ is, in fact, equal to $\dsR^3$ by construction. A point $u$ of $\mathcal{N}\dsR^2$ describes a vector $v$ normal to the plane $\dsR^2$ that is attached at a point $\mathbf{R}$ of $\dsR^2$. Its fiber coordinate is the vector $v$ and the base coordinate is the point $\mathbf{R}$. Parallel transport on $\mathcal{N}\dsR^2$ corresponds to moving normal vectors along a curve on the plane without rotating them or changing their lengths. The fiber coordinate does not change along the path upon parallel transport (Fig.~\ref{fig:parallel-transport}a), indicating that $\mathcal{N}\dsR^2$ is flat. The parallel transport of the fiber elements along any closed loop such as $\mathbf{R}_1 \mathbf{R}_2 \mathbf{R}_3 \mathbf{R}_4$ does not result in a holonomy.

Figure~\ref{fig:parallel-transport}b illustrates parallel transport on the tangent bundle of a hemisphere of unit radius. The tangent bundle is trivial because the hemispherical base manifold is equivalent to the Euclidean plane. Geometric considerations imply that parallel transport of a tangent vector preserves its length and does not rotate it about the local normal~\cite{bernstein1981fiber,shapere1989geometric}. For example, parallel transport of the vector $v_i$ over a \enquote{small} distance from $\mathbf{R}_1$ to $\mathbf{R}_2$ results in the vector $v_2$. Other possibilities, such as $v_2'$, are ruled out as they change the length of the vector or involve rotation about the local normal. The connection describing the parallel transport is curved~\cite{carmo_differential_2016}. Parallel transport of a vector around a closed loop, such as $\mathbf{R}_1 \mathbf{R}_2 \mathbf{R}_3 \mathbf{R}_4$, results in a holonomy. The holonomy is observed as an angle of rotation $\g$ of the tangent vector. The holonomy acquired over a loop $L$ is calculated using the Gauss-Bonnet theorem~\cite{carmo_differential_2016} as 
\begin{equation}
\label{eqn:classical-holonomy}
    \g(L) = \int_S \kappa \, \dd s\;,
\end{equation}
where $\kappa$ is the curvature of the base manifold, and $S$ is the surface encircled by the loop $L$. The hemisphere has unit curvature ($\kappa = 1$), and thus, the holonomy equals the solid angle subtended by the loop $L$. Perturbations to the loop $L$ continuously change the holonomy $\g$, suggesting a geometric nature of the holonomy. Identical considerations apply to the tangent bundle of a sphere $TS^2$.

The third example is parallel transport on a thickened M\"obius strip. A thickened M\"obius strip (Fig.~\ref{fig:parallel-transport}c) is constructed in analogy to the M\"obius strip but with an annular base manifold. It is a nontrivial fiber bundle. Parallel transport of a fiber element preserves its distance and orientation with respect to the center plane, implying a flat connection. Parallel transport over two representative loops $L_1$ and $L_2$ are considered. First, consider the loop $L_1$ encircling the center of the annulus. The fiber bundle restricted to this loop is a M\"obius strip shown in Fig.~\ref{fig:parallel-transport}d. Parallel transport of a fiber element $v_i$ around the M\"obius strip changes its sign, i.e., $v_i=-v_f$. Next, consider the loop $L_2$ of the base manifold not encircling the center of the annulus. Parallel transport of a fiber element on this loop is identical to parallel transport on the normal bundle $\mathcal{N}\dsR^2$ (Fig.~\ref{fig:parallel-transport}a). Consequently, no holonomy is observed. In general, the change in sign of the fiber element transported along a closed loop depends only on whether the loop encircles the center of the annulus. This indicates the topological nature of the holonomy.

The examples characterize the appearance of holonomy in fiber bundles with a connection. A trivial fiber bundle with a flat connection does not exhibit holonomy. A trivial fiber bundle with a curved connection exhibits holonomy with a geometric characteristic. The holonomy is calculated, in general, using analogs of the Gauss-Bonnet theorem~\cite{berry1984quantal}. A nontrivial fiber bundle with a flat connection exhibits holonomy with a topological characteristic. 

The above characterization explains the different mechanisms required to obtain a holonomy with either geometric or topological characteristics. Holonomy with a geometric nature depends on local features of the fiber bundle, namely the curvature induced by the connection, which can be computed pointwise~\cite{nakahara2003geometry}. Holonomy with a topological nature depends on a global feature of the fiber bundle, namely the nontrivial nature of the fiber bundle. Note that trivial and nontrivial fiber bundles are locally indistinguishable because local neighborhoods of any fiber bundle are equivalent to Euclidean spaces. In general, quantities related to the fiber bundle that continuously change with perturbations (geometric nature) depend on local features, and quantities that are invariant to perturbations (topological nature) depend on global features.

\section{Geometric phase as the holonomy of parallel transport}
\label{sec:geometric-phase-holonomy}
In view of the concepts introduced in the previous section, the geometric phase in adiabatically varying elastic waveguides (Sec.~\ref{sec:waves}) can be described mathematically by recasting the propagation of waves as parallel transport on a fiber bundle. However, the way to illustrate this concept depends on whether the geometric phase affects either the overall phase or the polarization angle. The two cases are treated separately, and the general conclusions will be drawn at the end of the section. 

\subsection{Effect on the overall phase}
\label{sec:dynamical-phase-holonomy}
\begin{figure}
    \centering
    \includegraphics[width=\linewidth]{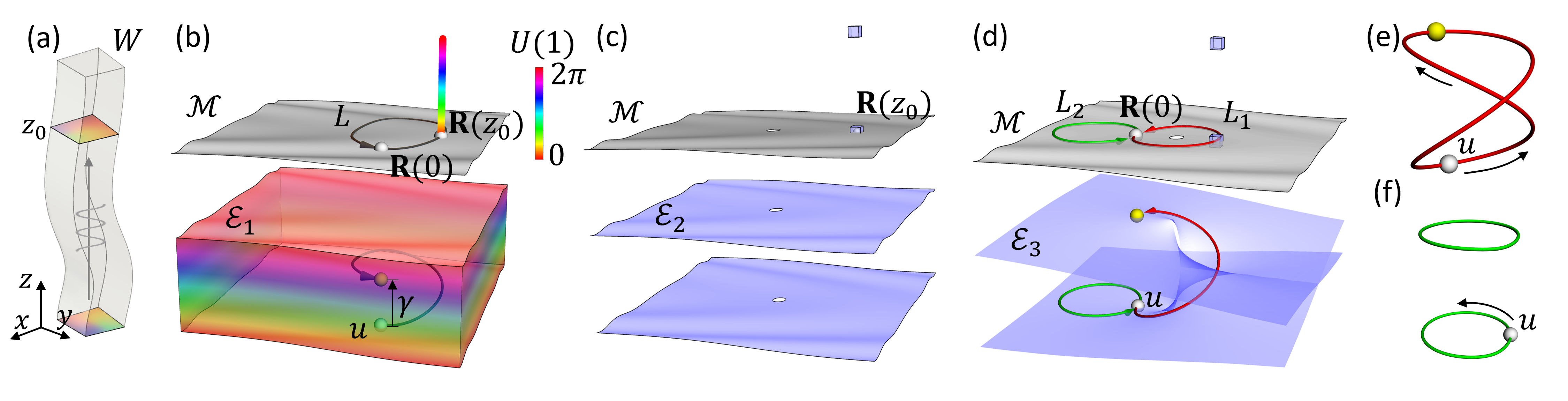}
    \caption{The geometric phase in an elastic waveguide is modeled using a fiber bundle. (a) A quasi-one-dimensional waveguide $W$ with cross sectional properties $\mathbf{R}$ varying in an adiabatic cycle along $z$. The waveguide is excited by a guided mode eigenfunction $\mathbf{U(x,y,\mathbf{R}(0))}$ at $z=0$. The disturbance propagates to $z=z_0$, where the response acquires a geometric phase $\g$. (b) Base manifold $\mathcal{M}=\dsR^2$ and fiber $U(1)$ resulting in the fiber bundle $\mathcal{E}_1$. Waveguide $W$ corresponds to the loop $L$ on the base manifold, which is parameterized by the $z$ coordinate. The fiber $U(1)$ is represented by unwrapping it into a line, where the color denotes the fiber coordinate. The fiber is attached at each point, as shown at $\mathbf{R}(z_0)$, to create the fiber bundle $\mathcal{E}_1$. The colors on the fiber bundle also correspond to the fiber coordinate. Parallel transport of $u$ along $L$ models wave propagation in $W$. It results in a holonomy $\g$, which is a nontopological geometric phase. (c) Base manifold $\mathcal{M}=\dsR^2-\{\mathbf{0}\}$ and fiber $\{0,\pi\}$ resulting in a trivial fiber bundle $\mathcal{E}_2$. The fiber $\{0,\pi\}$, depicted by blue cubes, is attached at a point $\mathbf{R}(z_0)$. (d) Base manifold $\mathcal{M}=\dsR^2-\{\mathbf{0}\}$ and fiber $\{0,\pi\}$ resulting in a nontrivial fiber bundle $\mathcal{E}_3$. Two loops $L_1$ and $L_2$ are marked. Parallel transport of an element $u$ along $L_1$ ($L_2$) results in a topological geometric phase (no geometric phase). (e) The edge of a M\"obius strip, obtained by restricting $\mathcal{E}_3$ to loop $L_1$. Parallel transport of $u$ is highlighted. (f) The edges of a cylinder, obtained from restricting $\mathcal{E}_3$ to loop $L_2$. Parallel transport of $u$ is highlighted.}
    \label{fig:waveguide-fiber-bundle}
\end{figure}

Consider a quasi-1D semi-infinite waveguide $W$ whose cross sectional properties are described by a set of $n$ parameters, $\mathbf{R}(z)$ (Fig.~\ref{fig:waveguide-fiber-bundle}a). The parameters vary adiabatically in a cycle along the longitudinal direction ($z$ axis). The space of the parameter values is generally equivalent to $\dsR^n$. Each parameter value $\mathbf{R}$ can define a uniform waveguide and its guided modes. Recall from Sec.~\ref{sec:waves} that a parameter value $\mathbf{R}_0$ resulting in a waveguide with fully degenerate modes is said to be a full degeneracy. The parameter space generally consists of nondegenerate values; only specific parameter values $P$ are fully degenerate. The nondegenerate parameter values defines the base manifold $\mathcal{M}$, that is, $\mathcal{M}=\dsR^n - P$ (Figs.~\ref{fig:waveguide-fiber-bundle}b-d). A parameter space without full degeneracies results in a Euclidean base manifold and leads to a trivial fiber bundle (Fig.~\ref{fig:waveguide-fiber-bundle}b). A parameter space with full degeneracies generally leads to a nontrivial fiber bundle (Fig.~\ref{fig:waveguide-fiber-bundle}d). 

The waveguide $W$ corresponds to a closed loop $L$ on the base manifold, which is parameterized by the longitudinal coordinate $z$ (Fig.~\ref{fig:waveguide-fiber-bundle}b). Let the waveguide be harmonically excited with angular frequency $\w$ triggering a flexural mode whose eigenfunction is $\mathbf{U}(x,y;\mathbf{R}(0))$ at $z=0$. The steady state displacement field at a location $z_0$ is determined up to the geometric phase as (Sec.~\ref{sec:waves})
\begin{equation}
    \mathbf{u}(x,y,z_0,t)=a(z) \mathbf{U} (x,y;\mathbf{R}(z_0)) \exp(\ii \g) \exp\left( -\ii \w t - \ii\int_0^{z_0} k(z') \dd z'\right) \;.
\end{equation}
The geometric phase $\g$ accounts for the phase ambiguity of the guided mode eigenfunction (Sec.~\ref{sec:discrete}), that is, if $\mathbf{U} (x,y;\mathbf{R}(z_0))$ is an eigenfunction, then so is $\mathbf{U} (x,y;\mathbf{R}(z_0)) e^{\ii \g}$. In general, the geometric phase $\g$ may assume any value from $0$ to $2\pi$, that is, any point on the complex unit circle $U(1)$. The geometric phase is described as the fiber coordinate on the fiber $U(1)$ attached to the point $\mathbf{R}$ of the base manifold. For ease of visualization, the fiber is unwrapped into a line whose color indicates the fiber coordinate (inset of Fig.~\ref{fig:waveguide-fiber-bundle}b). The base manifold and fibers give rise to a fiber bundle (Fig.~\ref{fig:waveguide-fiber-bundle}b). In systems with time reversal symmetry, such as conservative elastic waveguides, the guided modes are described by real-valued eigenfunctions. Thus, a computed eigenfunction is ambiguous only in its sign, and the geometric phase is restricted to $\{0,\pi\}$. The geometric phase is described by the fiber $\{0,\pi\}$. Some possible fiber bundles are shown in Figs.~\ref{fig:waveguide-fiber-bundle}c,d. In this manner, the response of a waveguide to a guided mode excitation has been modeled as a fiber bundle with base manifold as the space of nondegenerate parameters $\mathbf{R}$ and the fiber as $U(1)$ or $\{0,\pi\}$.

A connection on this fiber bundle is provided by the governing equations of motion and the adiabatic theorem. The adiabatic theorem limits the allowable magnitude of change in the guided mode eigenfunction over a small longitudinal distance $\Delta z$ to be of the order of $(\Delta z)^2$. That is,
\begin{equation}
    \langle \mathbf{U}(x,y;\mathbf{R}(z)) | \mathbf{U}(x,y;\mathbf{R}(z+\Delta z)) - \mathbf{U}(x,y;\mathbf{R}(z)) \rangle \sim \mathcal{O}((\Delta z)^2)\;,
\end{equation}
where the inner product $\langle \mathbf{f} | \mathbf{g} \rangle$ is an integral over the cross section $S$ defined as $\int_S \mathbf{f}^* \cdot \mathbf{g} \, \dd x \dd y$. Dividing the above equation by $\Delta z$ and imposing the limit $\Delta z \to 0$ provides
\begin{equation}
\label{eqn:berry-connection}
 \left \langle \mathbf{U}(x,y;\mathbf{R}(z)) \left | \frac{\partial \mathbf{U}(x,y;\mathbf{R}(z))}{\partial z}  \right. \right \rangle = 0\;.
\end{equation}
Equation~\eqref{eqn:berry-connection} defines a connection called the adiabatic connection or the Berry connection~\cite{berry1984quantal,simon1983holonomy}. If the fiber is $\{0,\pi\}$, the adiabatic connection simply ensures that the fiber coordinate does not change abruptly (from $0$ to $\pi$ or vice versa) as the base point moves in a neighborhood equivalent to the Euclidean space, hence implying that the connection is flat.

Having constructed the fiber bundle and identified a connection describing the dynamics of the waveguide $W$, the propagation of the excitation can be described as a parallel transport. This process is illustrated by Fig.~\ref{fig:waveguide-fiber-bundle}b. The initial cross section corresponds to a point $u$ on the fiber bundle with base coordinate $\mathbf{R}(0)$ and fiber coordinate zero, since the geometric phase at the initial value can be arbitrarily set to zero. As the wave propagates down the longitudinal coordinate $z$ in the adiabatically varying waveguide, the point on the base manifold travels on the loop $L$ parameterized by $z$ and the fiber coordinate is parallel transported according to the connection. The fiber coordinate at the end of the loop $L$, if nonzero, is a holonomy. However, given that the fiber coordinate is the geometric phase $\g$, then the geometric phase is the holonomy arising from parallel transport. 

If the fiber bundle is trivial with fiber $U(1)$, the geometric phase is nontopological, and it is computed as~\cite{berry1984quantal} 
\begin{equation}
\label{eqn:geometric-phase}
   \g(L) = \oint_L \mathrm{Im}\left[ \left\langle \mathbf{U}(x,y;\mathbf{R}(z)) \left| \frac{\partial \mathbf{U}(x,y;\mathbf{R}(z))}{\partial z} \right. \right\rangle \right] \cdot \dd \mathbf{R} \;.
\end{equation}
Examples of such geometric phase in elastic waves can be found in Refs.~\cite{budden1976phase,tromp1994surface}.  

If the fiber bundle is nontrivial with fiber $\{0,\pi\}$, the geometric phase is topological. This situation is exemplified by the discussions on flexural waves propagating in 1D waveguides with adiabatically-varying triangular cross section (Sec.~\ref{sec:1d-waveguide}). The waveguide property varying along the longitudinal direction is a geometric perturbation $\mathbf{R}=(\Delta x,\Delta y)$ to the cross section. The perturbation parameters lead to a two-dimensional parameter space with a full degeneracy at $\mathbf{R}=(0,0)$ arising from the symmetry of the corresponding uniform waveguide, which has an equilateral triangular cross section. Thus, the base manifold is the punctured plane $\dsR^2-\{\mathbf{0}\}$, and the fiber is $\{0,\pi\}$. 
The base manifold and fiber gives rise to one of the following fiber bundles: (i) a stack of two punctured planes $\mathcal{E}_2$ (formally $(\dsR^2-\{\mathbf{0}\}) \times \{0,\pi\}$; which is trivial) shown in Fig.~\ref{fig:waveguide-fiber-bundle}c, or (ii) a thickened M\"obius strip-like bundle $\mathcal{E}_3$ (nontrivial) visualized in Fig.~\ref{fig:waveguide-fiber-bundle}d (c.f. Sec.~\ref{sec:fiber-bundles}). Note that the self-intersection is an artifact of describing the fiber bundle in 3D space. As the simulations of Sec.~\ref{sec:1d-waveguide} confirm the appearance of a $\pi$-valued geometric phase, the fiber bundle must be nontrivial (because a trivial fiber bundle with flat connection does not exhibit holonomy). 

Figure~\ref{fig:waveguide-fiber-bundle}d explains the appearance of the topological geometric phase. A waveguide generated by parameters (not) encircling the full degeneracy corresponds to the loop ($L_2$) $L_1$ on the base manifold. This waveguide is described by a fiber bundle obtained as the restriction of $\mathcal{E}_3$ on ($L_2$) $L_1$. The resulting fiber bundle is the (top and bottom edges of a cylinder, Fig.~\ref{fig:waveguide-fiber-bundle}f) edge of a M\"obius strip, shown in Fig.~\ref{fig:waveguide-fiber-bundle}e. As an element $u$ of the fiber bundle with initial base coordinate $\mathbf{R}(0)$ and fiber coordinate zero is subject to parallel transport along ($L_2$) $L_1$, the fiber coordinate (remains zero for loop $L_2$) changes to $\pi$ for loop $L_1$. The topological geometric phase results from the topological difference between the edge of a M\"obius strip and the edges of a cylinder.

\subsection{Effect on the polarization angle}
\label{sec:polarization-angle-holonomy}
For this case, the construction of the fiber bundle and the identification of the connection is discussed using the example of polarized flexural waves propagating in a waveguide (Sec.~\ref{sec:helical-waveguides}).

Consider a waveguide $W$ generated by the closed loop $C_2$ on the sphere (Fig.~\ref{fig:polarization}g,h). Let the path coordinate along the centerline of the waveguide be $s$. Let the waveguide $W$ be excited by a flexural guided mode, assumed to be horizontally polarized. At any location $s$, the displacement field is determined up to the polarization vector $\Hat{\mathbf{c}}$ (Sec.~\ref{sec:waves}), which can be any unit vector in $\dsR^2$. It is modeled as the fiber coordinate on the fiber $\dsR^2$ attached to the point $\Hat{\mathbf{t}}$ of the sphere. The resulting fiber bundle is equivalent to the tangent bundle of the sphere $TS^2$ (Fig.~\ref{fig:manifold-FB}e).

The connection on the fiber bundle is imposed by the governing equations and the adiabatic theorem. Remarkably, the connection coincides with the natural connection on $TS^2$ arising from geometric considerations discussed in Sec.~\ref{sec:parallel-transport}, implying that it is curved~\cite{segert1987photon}. The polarization vector undergoes parallel transport on the spherical base manifold exactly like a tangent vector (Fig.~\ref{fig:parallel-transport}b). Thus, holonomy manifests as a change in polarization angle $\g$, which is the geometric phase. This term is calculated using Eq.~\eqref{eqn:classical-holonomy}.

\subsection{General considerations}
The previous two sections highlighted a few general considerations for a waveguide $W$ with adiabatically varying parameters $\mathbf{R}$ along the propagation coordinate $z$ (or $s$): (i) the propagation of waves resulting from a harmonic excitation is described as a parallel transport on a fiber bundle, (ii) the geometric phase is the holonomy arising from this parallel transport. The classification of holonomies according to the nature of the fiber bundle (trivial or nontrivial) and curvature of the connection (flat or curved) applies to geometric phases. In particular, a waveguide exhibiting a topological geometric phase has full degeneracies in the parameter space and is described by a nontrivial fiber bundle analogous to the edge of a M\"obius strip. A waveguide exhibiting a nontopological geometric phase is described by a fiber bundle with a curved connection. This classification is verified in the examples in Sec.~\ref{sec:waves}.  

\section{Topological classification of elastic waveguides}
\label{sec:applications}
The previous discussion on the geometric phase has illustrated the role of differential geometry and topology in understanding the dynamics of elastic waves. In this section, we show how the relation between geometric phase, topology, and dynamic response can be leveraged to design elastic waveguides.

In particular, we review the role of the geometric phase in the context of the design of elastic topological metamaterials (ETMs). Broadly speaking, elastic metamaterials~\cite{hussein2014dynamics,jiao2023mechanical} are typically periodic structures obtained by repeatedly assembling in space a section (called the unit cell) that embeds all the fundamental structural features of the material. The unit cell properties (e.g., geometry, materials, symmetries, etc.) can be tailored to enable a wide range of unconventional static and dynamic responses, such as negative Poisson's ratio~\cite{babaee20133d}, negative refractive index~\cite{yang_focusing_2004}, nonreciprocity~\cite{nassar_nonreciprocity_2020}, and cloaking~\cite{farhat_ultrabroadband_2009,stenger_experiments_2012}, just to name a few. Given the vastness of the design space, it is very challenging to comprehensively explore unit cell designs in search of specific material response. However, topology can be leveraged as an additional tool to support this search. Certain combinations of the unit cell properties, particularly those involving restrictions on spatial and temporal symmetries, lead to a specific class of metamaterials known as ETMs. The waveguiding properties of ETMs are closely connected to the topological properties of the dispersion relations, in a manner that is elucidated in the remainder of this section.

The ability to characterize the material behavior using topological concepts was first identified in the context of band structure analysis of condensed matter systems. The topological classification of abstract spaces and the physical classification of materials were unified in a remarkable chain of discoveries between 1970 and 2000 (see Refs.~\cite{hasan_colloquium_2010,chiu_classification_2016}). These studies showed how naturally occurring crystalline materials could be modeled as fiber bundles, and the corresponding physical properties were linked to topological invariants \footnote{Mathematical parameters that identify global properties of abstract mathematical spaces; in geometry this parameter is the genus. Topological invariants do not change under continuous deformations of the space. For example, a sphere has genus $g=0$. The sphere can be continuously morphed into a parallelepiped, which is also described by $g=0$; therefore the sphere and the parallelepiped are topologically equivalent objects.}. As an example, electric polarization was understood as the holonomy of parallel transport~\cite{resta_macroscopic_1994}. These investigations led to a \enquote{periodic table} of topological materials~\cite{chiu_classification_2016}. According to this classification scheme, materials can be categorized based on the topology of their band structure and on the value of the associated invariant. This invariant captures the global (topological) properties of the band structure, hence characterizing the dynamic behavior of the medium at a much deeper level that is independent from specific values of the material parameters. Materials characterized by the same invariant form a topological material class, which means that the topological nature of their band structure is intrinsically equivalent. On the other side, materials in distinct topological classes have inequivalent band structures that are typically separated by degeneracies; more precisely, if a variation of the material parameters changes the topological class, the corresponding band structure cannot be morphed into each other unless the band gap closes and reopens (i.e. degeneracies are created and annihilated). Further, in accordance to the bulk-boundary correspondence principle~\cite{hasan_colloquium_2010,chiu_classification_2016}, an interface between two topologically distinct materials, whose invariants differ by $n$ units, supports $n$ localized excitations called edge modes. These edge modes are protected against back-scattering produced by any potential defect or perturbation of the interface because the edge modes are determined by the global (topological) properties of the bulk, not by the local properties of the interface. The only way to alter the edge modes is to destroy the topological property of the bulk. These unique properties of the edge modes have led to technological advances in electronics and information processing~\cite{gilbert2021topological}.

The success of the \enquote{periodic table} of (quantum) topological materials spurred new directions such as higher-order topological metamaterials~\cite{xie_2021_Higherorder}, non-Abelian materials~\cite{yang_2024_NonAbelian}, and topological defects~\cite{lin_2023_Topological}. Each direction further built upon the connections between the topological characterization of the band structure and the corresponding exotic material behaviors. In each of these new material classes, a material is described as a fiber bundle and relates abstract properties of the fiber bundle to the observable behavior of the material.

In the last decades, researchers have successfully adapted some of these concepts to engineer classical (i.e. non-quantum) topological metamaterials for elastic waves, acoustics, and photonics~\cite{susstrunk_classification_2016,shah_2024_Colloquium,ozawa2019topological}. In the context of elastic topological metamaterials (ETMs), research focused on the identification and realization of material architectures capable of mimicking the dynamic behavior of their quantum mechanical counterparts. It is important to highlight the foundational difference between quantum and classical materials (whose behavior is ruled by quantum and classical mechanics, respectively) ultimately allows only analogies between the two material systems.

The first generation of ETMs replicated specific classes from the \enquote{periodic table} of quantum topological materials~\cite{susstrunk_classification_2016}. Analogous to their quantum counterparts, such ETMs mapped to fiber bundles and topological invariants, and interfaces between topologically distinct ETMs supported robust, localized edge modes according to the bulk-boundary correspondence. The desirable properties of the edge modes found technological applications. For example, in a 2D ETM, a line interface between topologically distinct ETMs supported localized propagating waves immune to back scattering even in the presence of inhomogeneities and defects~\cite{liuSyntheticKramersPair2021}. Soon after, ETMs based on higher-order topology~\cite{fan2019elastic} and topological defects appeared~\cite{chen2019mechanical}.

The initial development of ETMs was rapid since most designs emulated the design characteristics of their quantum counterparts. Nevertheless, researchers have only begun exploring ETM designs and capabilities~\cite{allein_strain_2023,zhang_second_2023}. To support the development of ETMs, design tools rooted in differential geometry and topology are critical. However, currently these ideas are spread out over different fields and making their role in the design process of ETMs not always directly accessible. This section presents an attempt to fill this gap by using the notions introduced in Sec.~\ref{sec:theory}. In particular, it highlights the link between waveguides, fiber bundles, topological invariants, and dynamics using simple (yet generalizable) examples.

\subsection{1D waveguides classified using topological geometric phase}
\begin{figure}
    \centering
    \includegraphics[width=\linewidth]{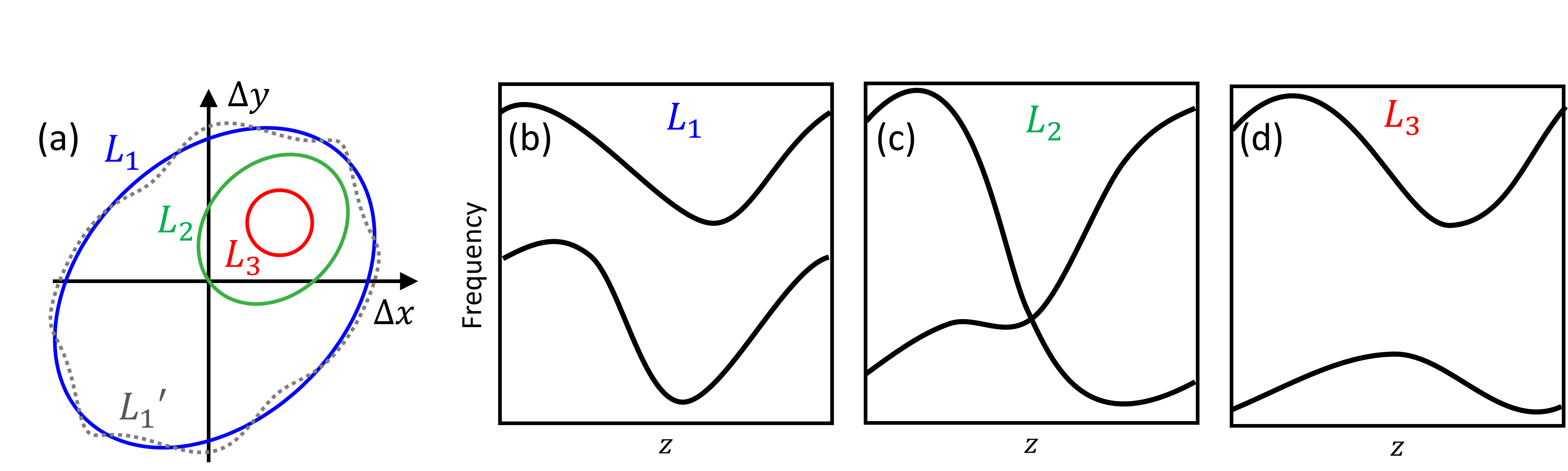}
    \caption{(a) Parameter space used to generate the 1D waveguides with adiabatically varying triangular cross sections. Each loop corresponds to a waveguide. (b), (c), (d) show the $\w$-$z$ plots corresponding to loops $L_1$, $L_2$, and $L_3$ at a fixed $k$.}
    \label{fig:waveguide-classes}
\end{figure}

Consider the adiabatically-varying triangular waveguides presented in Sec.~\ref{sec:1d-waveguide}. A waveguide $W$ can be visualized as a closed loop, say $L_1$, on the base manifold $\dsR^2-\{\mathbf{0}\}$ (Fig.~\ref{fig:waveguide-classes}a), as discussed earlier. $L_1$ is parameterized by the longitudinal coordinate $z$.
A waveguide can also be visualized in terms of the dynamical properties of its cross section. Each point $\mathbf{R}$ on the loop $L_1$ corresponds to a cross section of the waveguide $W$ at some coordinate $z$ along the longitudinal axis. This cross section, with parameters $\mathbf{R}(z)$, defines a uniform waveguide denoted $W_{\mathbf{R}(z)}$. The dynamical properties of the flexural guided modes of the uniform waveguide $W_{\mathbf{R}(z)}$ are described by the dispersion relations $\w(k,\mathbf{R}(z))$. By allowing $z$ to vary, $\w(k,\mathbf{R}(z))$ can capture the dispersion relations of every cross section of $W$. For a convenient visualization, we fix $k$ and plot $\w$ as a function of $z$ in Fig.~\ref{fig:waveguide-classes}b-d. This plot is referred in the following as the `$\w$-$z$' plot. The two curves in a $\w$-$z$ plot generally do not intersect. The curves intersect if, and only if, the flexural guided modes for some $W_{\mathbf{R}(z)}$ are fully degenerate, which means that the loop corresponding to $W$ passes through the full degeneracy (the origin). This situation is illustrated by loop $L_2$ in Fig.~\ref{fig:waveguide-classes}a and the corresponding $\w$-$z$ plot in Fig.~\ref{fig:waveguide-classes}c. We digress to note that while the point degeneracy on plots analogous to $\w$-$z$ curves is crucial in the context of topological classification, its connection with a full degeneracy on the dispersion relation is a peculiarity in the present system. In general, full degeneracies are rare in the general context of topological classification.

The first perspective establishes a classification scheme of the waveguides into two types, $\scW_O$ and $\scW_E$. Let the number of times the loop corresponding to a waveguide encircles the origin be denoted by $\nu$. Then, $\scW_O$ ($\scW_E$) is the class of the waveguide for which $\nu$ is odd (even). Waveguides in class $\scW_O$ ($\scW_E$) acquire (do not acquire) a geometric phase of $\pi$. For example, in Fig.~\ref{fig:waveguide-classes}a, the waveguide corresponding to loop $L_1$ ($L_2$) is in the class $\scW_O$ ($\scW_E$), which acquires (does not acquire) a geometric phase of $\pi$.   

The properties of this classification scheme parallel the topological classification of metamaterials. Firstly, the classification scheme is robust against continuous geometric perturbations to a waveguide. Geometric perturbations to a given waveguide manifest as changes to the corresponding loop. If these perturbations are small enough, such as perturbing loop $L_1$ to $L_1'$ in Fig.~\ref{fig:waveguide-classes}a, the waveguide class does not change. Secondly, waveguides in different classes are separated by gap closures in the $\w$-$z$ plot. If a perturbation deforms a waveguide into a different class, there is an intermediate waveguide whose loop passes through the origin. For example, in Fig.~\ref{fig:waveguide-classes}a, perturbing the waveguide corresponding to loop $L_1$ into a waveguide corresponding to loop $L_3$ passes through an intermediate waveguide corresponding to loop $L_2$, for which the gap is closed in the $\w$-$z$ plot. 

Further, this classification scheme is also topological in nature. Recall from Sec.~\ref{sec:dynamical-phase-holonomy} that waveguides in class $\scW_O$ ($\scW_E$) give rise to a nontrivial (trivial) fiber bundle resembling the edge of a M\"obius strip shown in Fig.~\ref{fig:waveguide-fiber-bundle}e (edges of a cylinder, Fig.~\ref{fig:waveguide-fiber-bundle}f). Thus, the classification scheme essentially distinguishes between these trivial and nontrivial fiber bundles. The topological invariant is the holonomy of the parallel transport along the base manifold, which is precisely the topological geometric phase (Figs.~\ref{fig:waveguide-fiber-bundle}e,f).

\subsection{Topological classification of elastic metamaterials}
\begin{figure}
    \centering
    \includegraphics[width=\linewidth]{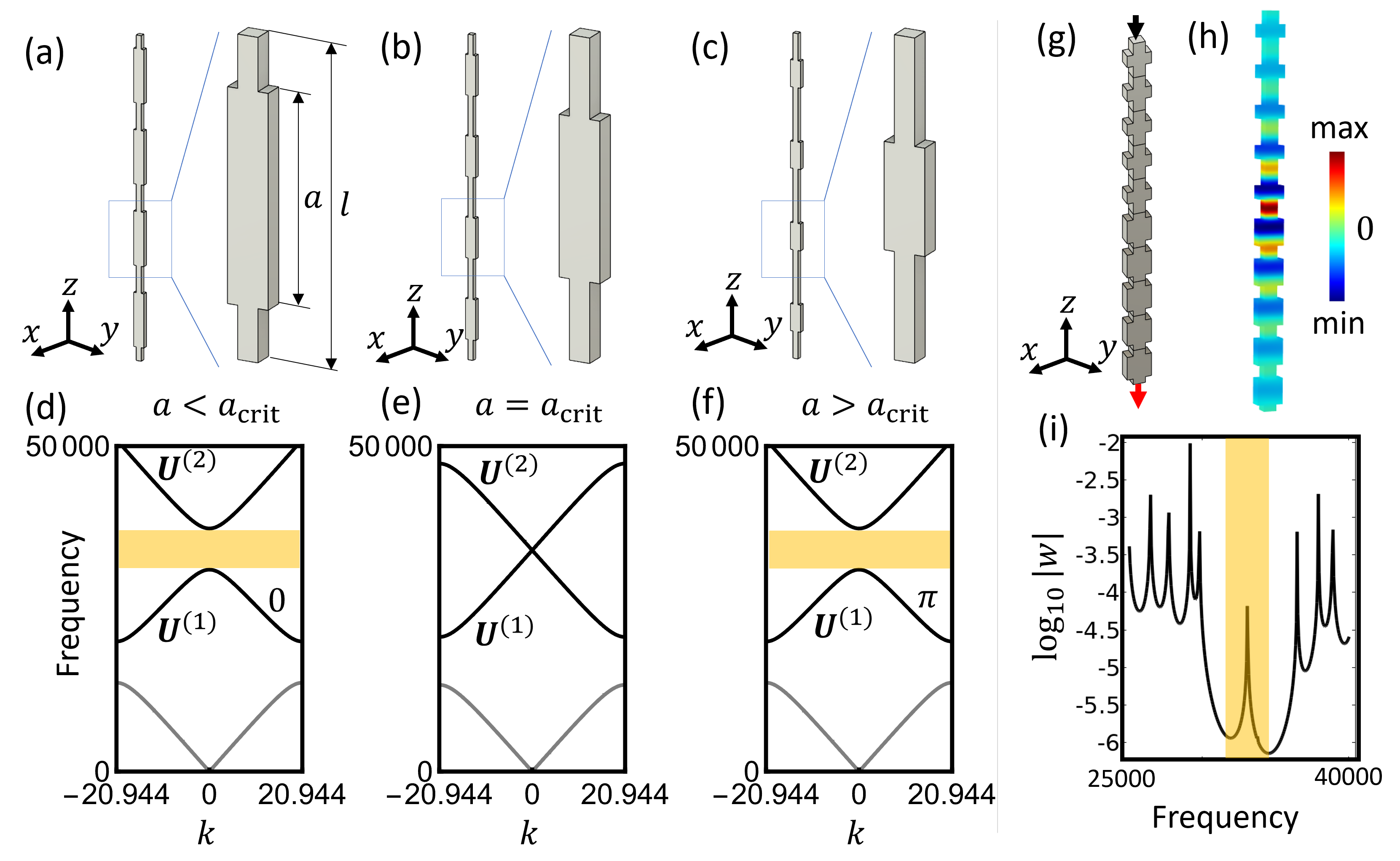}
    \caption{1D stepped beam waveguides with unit cell of length $l=15\mathrm{cm}$ of which the length $a$ is the thicker region. (a) $a=5\mathrm{cm}$, (b) $a=7.5\mathrm{cm}$, (c) $a=10\mathrm{cm}$. (d)-(f) Dispersion relations corresponding to the waveguides in (a)-(c) with the bandgaps highlighted. (g) A finite beam constructed by joining 5 unit cells with $a=5\mathrm{cm}$ and 5 units cells with $a=10\mathrm{cm}$. (h) The mass normalized $z$-component of displacement of a mode of vibration that is localized at the interface of the topologically distinct regions. The longitudinal $z$ coordinate has been scaled down by a factor of $5$ in (g) and (h) for better visualization of the waveguides. (i) Frequency response curve of the finite beam. The beam is subjected to harmonic excitation along the $z$ direction at the top end (black arrow in (g)) and the $z$ component of displacement response is measured at the bottom end (red arrow in (g)). The natural frequency of the localized mode is within the common frequency gap of the dispersion relations in (d) and (f). Figure adapted from Ref.~\cite{yin2018band}.}
    \label{fig:yin2018}
\end{figure}

The above classification scheme is readily adapted to ETMs once the fiber bundle description is identified. 1D ETMs that are infinite and periodic along the $z$ direction are considered. Further, the unit cell of length $a$ is assumed to be symmetric about the $xz$ and $yz$ planes to prevent coupling between flexural, torsional, and longitudinal modes. Such metamaterials admit solutions of the form $\mathbf{u}(x,y,z,t)=\mathbf{U}(x,y,z;k)e^{\ii(k z-\w t)}$ to the NL equations, where $\mathbf{U}(x,y,z;k)$ is a periodic function of $z$ called the Bloch wave mode~\cite{hussein2014dynamics,miniaci_design_2021}. The Bloch wave mode $\mathbf{U}(x,y,z;k)$ and angular frequency $\w(k)$ are computed by solving a wavenumber dependent eigenvalue problem. These properties of the ETM are graphically described by plotting the dispersion relation, which shows the angular frequency as a function of the wavenumber, e.g., Figs.~\ref{fig:yin2018}d-f. Due to the spatial periodicity of the metamaterial, the solutions are periodic in $k$ with periodicity $2\pi/a$, i.e., $\mathbf{U}(x,y,z;k)=\mathbf{U}(x,y,z;k+2\pi/a)$ and $\w(k)=\w(k+2\pi/a)$. In other words, the wavenumber is a circle $S^1$ represented as the interval $[-\pi/a,\pi/a]$ with the endpoints glued together. The wavenumber defines a circular base manifold. At each wavenumber $k$, the Bloch wave mode $\mathbf{U}(x,y,z;k)$ has a phase degree of freedom, as it can be arbitrarily multiplied by a phase factor $e^{\ii \theta}$. This is described as a $U(1)$ fiber. Thus, a Bloch wave mode of an ETM is naturally described as a fiber bundle.

Analogies from the previous section can now be exploited. The wavenumber $k$ plays the role of the longitudinal coordinate $z$. The dispersion relation is analogous to the $\w$-$z$ plot. Two specified Bloch modes $\mathbf{U}^{(1)}$ and $\mathbf{U}^{(2)}$ (e.g., corresponding to the longitudinal mode) correspond to the two curves of the $\w$-$z$ plot. A physical classification scheme can be established using the following criterion: ``Two waveguides belong to the same material class if they can be transformed into one another by continuous variation of either the geometric or the material properties without closing the gap between $\mathbf{U}^{(1)}$ and $\mathbf{U}^{(2)}$ in the dispersion relation''. This classification is also a topological classification of the fiber bundles. In this example, the topological invariant is the holonomy of parallel transport of a Bloch wave mode around the circular base manifold according to the Berry connection (Eq.~\eqref{eqn:berry-connection} with $k$ replacing $z$). It is called the Zak phase, computed using Eq.~\eqref{eqn:geometric-phase} as
\begin{equation}
\label{eqn:zak-phase}
    \theta_{\mathrm{Zak}} = \mathrm{Im} \left[ \int_{-\pi/a}^{\pi/a} \left\langle \mathbf{v} \left| \frac{\dd \mathbf{v}}{\dd k} \right. \right\rangle \dd k \right]\;.
\end{equation}
The Zak phase assumes a value of $0$ or $\pi$ modulo $2\pi$. Intuitively, the Zak phase distinguishes between twisted M\"obius strip-like fiber bundles and untwisted cylinder-like fiber bundles.

A concrete realization of the above ideas for longitudinal modes was studied by Yin et al.~\cite{yin2018band}. The results are reproduced in Fig.~\ref{fig:yin2018}. They considered stepped beams with a unit cell of fixed length $l=$~15~$cm$, as shown in Figs.~\ref{fig:yin2018}a-c. The effect of varying the length $a$ of the thicker portion of the unit cell on the dispersion relation was investigated. As $a$ was increased from $0$ to $l$, the geometry of the waveguide and its dispersion relation were plotted (see Figs.~\ref{fig:yin2018}a-f). It is seen from these results that the band gap between two Bloch modes $\mathbf{U}^{(1)}$ and $\mathbf{U}^{(2)}$ closes at a value $a=a_\mathrm{crit}=$~7.5~$cm$ (Fig.~\ref{fig:yin2018}e). The parameter value $a_{\mathrm{crit}}$ and $k=0$ results in an accidental degeneracy. This indicates that waveguides generated with $a<a_\mathrm{crit}$ and $a>a_\mathrm{crit}$ belong to potentially different material classes. The topological nature of the classification is verified by computing the Zak phase corresponding to the dispersion curve labeled as $\mathbf{U}^{(1)}$, which is $0$ ($\pi$) for waveguides with $a<a_\mathrm{crit}$ ($a>a_\mathrm{crit}$). 

The physical significance of the topological invariant manifests in the finite structure shown in Fig.~\ref{fig:yin2018}g created by joining two waveguides with Zak phases equal to $0$ and $\pi$, respectively. According to the bulk-boundary correspondence, the difference between the invariants of the constituent waveguides equals the number of localized edge modes at their interface. Here, the invariants differ by one unit because the Zak phase can only take values $0$ and $\pi$. Thus, the resulting structure possesses a single edge mode at the interface (Fig.~\ref{fig:yin2018}h) at a frequency within the common band gap of the two halves of the composite waveguide (Fig.~\ref{fig:yin2018}i). The mode is robust against small geometric or material perturbations that preserve the reflection symmetries of the structure as such perturbations do not change the topological class of the constituent waveguides.

In general, for ETMs emulating the \enquote{periodic table} of topological metamaterials, the ETM maps to a fiber bundle with base manifold as the Brillouin zone (circle for 1D ETM, 2-torus for 2D ETM, 3-torus for 3D ETM) and fiber as the Bloch wave mode. The topological invariant is computed using standard formulae derived in the literature~\cite{chiu_classification_2016}. Note that the invariant is, in general, not a geometric phase. The invariant is determined by the dimension of the waveguide and its symmetries. The topological invariant relates to edge modes at the interface of topological distinct ETMs through the bulk-boundary correspondence.

ETMs based on more recent design principles such as higher-order topological metamaterials or topological defects also benefit from these concepts. Indeed, higher-order topological metamaterials are generalizations of the 1D Zak phase~\cite{xie_2021_Higherorder,benalcazar_2017_Quantized} and phenomena at topological defects are extensions of the \enquote{periodic table} of topological metamaterials~\cite{lin_2023_Topological,chiu_classification_2016}.

\section{Conclusions}
\label{sec:conclusions}
This article reviewed the concept of geometric phase in the context of elastic waves and presented some applications to the design of dynamic and topological elastic materials. The geometric phase in quasi-1D waveguides was examined via practical examples, theoretical explanations, and design applications. The geometric phase was observed to depend solely on the path traced out in parameter space, and not on the rate of change of parameters. It can also be classified as either topological or nontopological. Only the former is robust against small perturbations to the design of the waveguide. The geometric phase was also discussed using the concepts of fiber bundles and parallel transport, hence leading to its interpretation as the holonomy resulting from parallel transport. Similarly, the topological nature was distinguished based on the nature of the fiber bundle and its connection. Topological geometric phases require nontrivial fiber bundles resembling the edge of a M\"obius strip, which are associated with full degeneracies in the parameter space describing the waveguide. Nontopological geometric phases require curved connections, which arise in rotating systems and in systems supporting polarized waves. The latter class of systems are described by the tangent bundle of a sphere. The interpretation of the geometric phase via arguments of differential geometry also enables its application to the design of ETMs. Indeed, ETMs can be described as fiber bundles classified according to topological invariants. Crucially, the abstract classification of fiber bundles indicates that the interface of ETMs with differing topological invariants can support localized edge modes. These edge modes are topologically protected against certain perturbations to the properties of the waveguide. 

The concept of geometric phase provides a rigorous and powerful tool to support either the theoretical understanding of ETMs or the design of novel ETMs. Indeed, the geometric phase complements other existing analysis tools to enable the selective exploration of the vast metamaterial design space in search of materials with unique properties. 

\enlargethispage{20pt}

\section*{Acknowledgements}
The authors gratefully acknowledge the partial financial support of the National Science Foundation under grant \#2330957. M.K. thanks Prof. Ralph Kaufmann for useful discussions. M.K. acknowledges financial support from the Ross fellowship administered by the Purdue University Graduate School.


\bibliographystyle{unsrt}
\bibliography{References}

\end{document}